\documentclass[tradictabstract]{aa}  
\usepackage{graphicx}
\usepackage{txfonts}

\newcommand{\irc}{IRC\,+\,10$^{\circ}$216}

\newcommand{\kms}{\mbox{km~s$^{-1}$}}

\newcommand{\mloss}{\mbox{$\dot{M}$}}
\newcommand{\my}{\mbox{$M_{\odot}$~yr$^{-1}$}}

\newcommand{\rstar}{\mbox{$R_{*}$}}

\newcommand{\vinf}{\mbox{$v_{\infty}$}}

\newcommand{\arcsecp}{\mbox{\rlap{.}$''$}} 
\newcommand{\secp}{\mbox{\rlap{.}$^{\mathrm{s}}$}}

\begin{document}

   \title{\irc\ mass loss properties through the study of $\lambda$3\,mm emission\thanks{Reduced data cubes are available in FITS format at the CDS via anonymous ftp to cdsarc.u-strasbg.fr (130.79.128.5) or via http://cdsweb.u-strasbg.fr/cgi-bin/qcat?J/A+A/}}

   \subtitle{Large spatial scale distribution of SiO, SiS, and CS}

   \author{L. Velilla-Prieto
          \inst{1,2}
          \and
          J. Cernicharo\inst{2}
          \and
          M. Ag\'undez\inst{2}
          \and
          J.\,P. Fonfr\'ia\inst{2}
          \and
          G. Quintana-Lacaci\inst{2}
          \and
          N. Marcelino\inst{2}
          \and
          A. Castro-Carrizo\inst{3}
          }

   \institute{Dpt. of Space, Earth and Environment, Chalmers University of Technology, Onsala Space Observatory, 439 92 Onsala, Sweden\\
              \email{luis.velilla@chalmers.se}
         \and
             Instituto de F\'isica Fundamental (IFF--CSIC), Serrano 123, Madrid, CP 28006 , Spain \\
         \and
            Institut de Radioastronomie Millimétrique, 300 Rue de la Piscine, F-38406 Saint-Martin d'Hères, France
         }

   \date{Received ; accepted }

    \abstract{Low-mass evolved stars are major contributors to interstellar medium enrichment
    as a consequence of the intense mass-loss process these stars experience at the end of their lives.
    The study of the gas in the envelopes surrounding asymptotic giant branch (AGB) stars through observations in the millimetre wavelength range provides 
    information about the history and nature of these molecular factories.
    Here we present ALMA observations at subarsecond resolution, complemented with IRAM-30\,m data, of several lines of SiO, SiS, 
    and CS towards the best-studied AGB circumstellar envelope, \irc.
    We aim to characterise their spatial distribution and determine their fractional abundances
    mainly through radiative transfer and chemical modelling.
    The three species display extended emission with several enhanced emission shells.
    CS displays the most extended distribution reaching distances up to approximately 20\arcsec. 
    SiS and SiO emission have similar sizes of approximately 11\arcsec, but SiS emission is slightly more compact. 
    We have estimated fractional abundances relative to H$_2$, which on average are equal to 
    f(SiO)$\sim$10$^{-7}$, f(SiS)$\sim$10$^{-6}$, and f(CS)$\sim$10$^{-6}$ up to the photo-dissociation region.
    The observations and analysis presented here show evidence that the circumstellar material displays clear deviations from 
    an homogeneous spherical wind, with clumps and low density shells that may allow UV photons from the interstellar medium (ISM) to 
    penetrate deep into the envelope, shifting the photo-dissociation radius inwards.
    Our chemical model predicts photo-dissociation radii compatible with those derived from the observations, 
    although it is unable to predict abundance variations from the starting radius of the calculations ($\sim$10\,\rstar), which may reflect the simplicity of the model.
    We conclude that the spatial distribution of the gas proves the episodic and variable nature of the mass loss mechanism of \irc,
    on timescales of hundreds of years.
    }

   \keywords{circumstellar matter ---
            stars: AGB and post-AGB ---
            stars: individual (\irc)
               }

   \maketitle

\section{Introduction} \label{sec:intro}
Circumstellar envelopes (CSEs) around low to intermediate mass stars are some of the most active sites of molecular and dust formation in the Galaxy. 
They have a major impact on synthesising material and enriching the interstellar medium (ISM) for subsequent galactic evolution 
\citep[see][for a general review on asymptotic giant branch (AGB) stars]{hab03}.
This has stimulated a large number of studies aimed at searching for and characterising the molecular emission of AGB stars 
since the opening of the (sub-)millimetre wavelength window \citep[e.g.][]{sol71,tur71,olo82,cer86}.
But as in other different domains of the electromagnetic spectrum, the limited spatial resolution achieved by single-dish radio telescopes
prevented astronomers from drawing solid conclusions about relevant questions such as 
AGB wind and mass-loss nature or about the formation processes of dust particles and molecules \citep[e.g.][]{cer89,vel17}.
Here we provide new observational constraints to the molecular emission of \irc\ impinging on 
the nature of its mass-loss history and photo-dissociation processes, 
thanks to the spatial resolution  and excellent sensitivity provided by the Atacama Large Millimetre/submillimetre Array (ALMA).

After decades of research and intensive observational efforts, \irc\ is still probably one of the most attractive objects for molecular study in the context of AGB CSEs.
CW\,Leo is the central star of the object, classified as the 216$^{th}$ object in the +10$^{\circ}$ declination band of the infrared catalogue \citep{neu69}.
It is located at $\sim$120\,pc away from us \citep{gro12} and its molecular envelope has been formed as a result of intense mass loss \citep[2--4$\times$10$^{-5}$\,\my,][and references therein]{gue18} of this 
C-rich Mira variable \citep[period$\sim$630\,days,][]{men12}.
The systemic velocity of the source with respect to the local standard of rest (LSR), $v_\mathrm{*}$, is -26.5$\pm$0.3\,\kms\ 
and the terminal expansion velocity of the circumstellar gas ($v_\mathrm{\infty}$) is 14.5$\pm$0.2\,\kms\ \citep{cer00}.

In terms of molecular content, it is the richest AGB CSE with a total of $\sim$80 different species detected (along with many other of their isotopologues), representing almost half of the total molecules found in the interstellar and circumstellar medium. 
\irc\ is also a prime target for studies of carbon dust formation in AGBs \citep{mar87,gro97,cer00,cer17}.
Several diatomic molecules are efficiently formed in the atmosphere of the star, such as CO, CS, or SiS,
where the chemistry is controlled mainly by thermodynamical equilibrium (TE) as well as by shocks due to the pulsation 
of the star \citep[][and references therein]{agu06,fon08,agu12,vel15}.
In particular, SiS and CS are predicted to be very abundant in the inner layers of C-rich AGB CSEs by TE models \citep[with average values of $\sim$10$^{-6}$ relative to H$_2$,][]{agu12}, 
and even higher in the case of SiS if the effect of periodic shocks are considered in the models \citep{che12}.
The case of SiO is particularly interesting, since most oxygen in C-rich stars is thought to be locked into the stable molecule CO while it is detected with significant abundances in the inner region of \irc\ 
\citep{agu12,fon14}.
TE models predict almost negligible abundances for SiO in the atmospheres of these C-stars, while higher abundances (up to $\sim$10$^{-5}$) are predicted at distances of $\sim$5\,\rstar\ \citep{agu12}.
Shock-induced chemistry due to the pulsation of the star may also play a role in the abundance of SiO and other molecules in the dust-formation region \citep{che12}.
Other scenarios may also contribute to the production of O-bearing species in the CSEs of C-rich stars and vice versa, such as an enhanced photo-induced chemistry due to clumpiness, 
as explored by \cite{agu10},
or \cite{van19}, or additionally if the extra source of UV radiation is a hotter binary companion.
SiS and SiO, as refractory molecules for the temperatures found in the region between 1 to 20\,\rstar, should have an important role in the condensation and growth of dust grains 
\citep[see e.g.][and references therein]{gai86,gai13a,gai13b}. 

From previous observational studies of rotational lines, including highly excited vibrational states, it is concluded that CS and SiS have typical abundances of $\sim$3$\times$10$^{-6}$ in the 
innermost layers of \irc\ decreasing down to $\sim$10$^{-6}$ at 10--20\,\rstar\ and then disappearing from the gas phase at $\sim$600\,\rstar\ in the case of SiS and at $\sim$1000\,\rstar\ for 
CS \citep{agu12}.
These authors also derived a constant SiO abundance of $\sim$10$^{-7}$ from the inner layers up to distances of $\sim$500\,\rstar.
The spatial distribution of these three molecules has also been traced with interferometric observations, showing roughly compact distributions extended up to $\sim$10--12\arcsec, equivalent to 
$\sim$520--630\,\rstar\ for SiS, $\sim$350--520\,\rstar\ for SiO, and $\sim$850--1000\,\rstar\ for CS \citep{luc92,luc95,bie93,pat11}.
These observations were obtained with a spatial resolution of $\sim$3--5\arcsec\ with the Institut de Radioastronomie Millim\'{e}trique (IRAM) Plateau de Bure Interferometer (PdBI),
the Berkeley-Illinois-Maryland Array (BIMA), and the Submillimetre Array (SMA).
It is worth noting that due to the lack of short-spacing observations the sizes listed before may be strictly considered as lower limits,
although this should have a limited impact in the case of compact distributions \citep[see the discussion for the case of SiS in][]{bie93}. 
Large-scale flux is provided by short-space observations, thus, single-dish observations must be considered if the aim is to study the complete molecular emission of an extended object.

In this paper we present our latest results from the study of several emission lines of SiS, SiO, and CS isotopologues detected in the 3\,mm wavelength spectrum of \irc\ using ALMA at 
subarsecond spatial resolution.
The observations and reduction methods are described in Section\,\ref{sec:obs}.
The observational results are presented in Section\,\ref{sec:res}.
In Section\,\ref{sec:mod} we present radiative transfer models aimed at deriving the radial abundance distribution of the three molecules studied. 
In Section\,\ref{sec:chem} we present an updated analysis of the chemistry of the studied species.
A consolidating discussion of the results is presented in Section\,\ref{sec:dis}.
Finally, in Section\,\ref{sec:conc} we present the conclusions of this work.

\section{Observations} \label{sec:obs}
The data presented here belong to observations completed with the ALMA and the IRAM-30\,m telescopes.
We carried out ALMA observations in band 3, corresponding to a $\lambda$\,3\,mm spectral line survey of \irc\ covering the frequency range $\sim$84.0-115.5 GHz during Cycle\,2.
The total frequency coverage was achieved by using five different spectral setups with an effective spectral resolution of 0.488\,MHz (channel spacing 0.244\,MHz).
We used the C34-7 and C34-2 configurations (extended and compact, respectively).
The extended configuration encompassed baselines up to $\sim$1600\,m, resulting in synthesised beam sizes as small as $\sim$0\arcsecp6.
The compact configuration provided baselines as low as 13\,m and up to 390\,m, resulting into maximum recoverable scales
in the range of $\sim$23\arcsec\ up to $\sim$32\arcsec\ depending on the frequency.
The observations were taken during several runs on December 4, 2014, and on June 29, July 4, 5, and 19, 2015. 
We did an initial calibration of the ALMA data with the Common Astronomy Software Applications \citep[CASA,][]{mcm07} by using the scripts provided by the ALMA team. 
To calibrate the gain (for phases and amplitudes) J0854+2006 or J0854+201 were observed. 
The quasar J1058+0133 was consistently used for the bandpass calibration. 
In half of the tracks, observations of Ganymede or Callisto were performed to calibrate absolute amplitudes, as their Butler-JPL-Horizons-2012 models were integrated in CASA and used as reference. 
For the other tracks, fluxes stored in the ALMA database for other calibrators (J0854+201, J1058+015, and J0854+2006) were used. 
From the differences obtained at the end for the fluxes of different calibrators and dates, we estimate an accuracy in the flux calibration better than 15\%.  
Finally, for every track the standard phase calibration was improved by self-calibrating with the very compact source continuum emission, by using the Grenoble Image and Line Data Analysis System (GILDAS) software package.
A single pointing was done towards the source position, with coordinates J2000.0 R.A.=9$^{\mathrm{h}}$47$^{\mathrm{m}}$57\secp446 and Dec.=13$^{\circ}$16$^{'}$43\arcsecp86
according to the position of the $\lambda$\,1\,mm continuum emission peak \citep{cer13}.
A complete description of these ALMA observations will be presented elsewhere (Cernicharo, in prep.).

IRAM-30\,m on-the-fly (OTF) observations were carried out on June 7--11, 2015, in order to recover the flux filtered out by ALMA.
With this purpose, we obtained OTF maps of 96\arcsec$\times$96\arcsec\ in the frequency range of $\sim$80--116\,GHz, with an average nominal
sensitivity of 5\,mK and a velocity resolution of 1.5\,\kms.
These maps were observed in position-switching mode with a position throw of $\sim$240\arcsec\ off-source.
Standard procedures of calibration, including pointing and focus on strong and nearby sources, were followed regularly during the observing runs.

Final data cubes were produced by merging the ALMA and IRAM-30\,m data
after continuum subtraction. This was done by using the UVSHORT task
in the GILDAS software package that uses the pseudo-visibility
technique to produce visibilities from single-dish maps, merge them to
interferometric visibilities, and use adapted deconvolution algorithms
to properly image the emission detected at several spatial scales
\citep[see][and references therein for more details]{pet10}.  
The relative amplitude calibration between ALMA and 30m OTF
data was particularly checked in the {\it uv}-plane, in the interface
between the different visibilities, before image synthesis.

Two final data cubes were generated for every spectral window. The first
was produced with a lower spatial resolution by using the natural visibility
weights. In this case, we obtain optimum point-source sensitivity with
moderated spatial resolution.  A second data cube was obtained by
using a robustness factor of 0.5 during the deconvolution process to
adapt the visibility weights in order to reach a higher angular resolution, though
with a slight decrease in sensitivity. In both cases, the
Steer-Dewdney-Ito algorithm \citep[SDI,][]{ste84} was used during
the cleaning process, given that \citet{hog74} can
introduce artificial clumpiness when generating images of objects that have extended 
emission. In all the cases a comparison was made between
the different cleaning methods to ensure the consistency and quality 
of our results. A list of the emission lines observed and the corresponding synthesised beams is given in Table\,\ref{tab:summary}.

\begin{table*}[hbtp!]
\caption{Summary of parameters for observed lines.}
\label{tab:summary}
\begin{center}
\begin{tabular}{l c c c c c c}
\hline\hline
Line                                    & Rest Frequency & $\theta_{\mathrm{s}}$        & PA                                      & $\sigma$                                      &         S$_\mathrm{\nu,max}$    & M$_\mathrm{0,max}$    \\
-                                       & (MHz)                         & (\arcsec)                                       & ($^{\circ}$)  &       (mJy\,beam$^{-1}$)      &       (Jy\,beam$^{-1}$)          & (Jy\,beam$^{-1}$\,\kms)      \\
\hline                                                                                                           \\ 
SiO $J$=2--1                    &  86846.986            & 0.60$\times$0.53                         &       41                                      &       1                                                                 &  0.14                                                         & 1.8      \\ 
$^{29}$SiO $J$=2--1 &  85759.194        & 0.86$\times$0.72                      & 210                             &       1                                                               &       0.02                                                            & 0.3      \\ 
$^{30}$SiO $J$=2--1 &  84746.165        & 0.87$\times$0.73                      & 210                             &       1                                                               &       0.02                                                            & 0.14     \\ 
SiS $J$=5--4                    &  90771.566            & 0.79$\times$0.60                         &       39                                      &       2                                         & 1.8                                                           &11.3    \\ 
$^{29}$SiS $J$=5--4 &  89103.750        & 1.14$\times$0.87                      &       51                                      &       2                                                                & 0.03                                                          & 0.5      \\ 
$^{30}$SiS $J$=5--4 &  87550.559        & 0.81$\times$0.66                      &       26                                      &       1                                                                 & 0.13                                                          & 0.2      \\ 
Si$^{34}$S $J$=5--4 &  88285.830        & 1.14$\times$0.88                      &       51                                      &       1                                                                 & 0.02                                                          & 0.4      \\ 
SiS $J$=6--5                    & 108924.303            & 0.79$\times$0.70                         &       57                                      &       2                                                                 &       0.4                                                             & 9.2     \\ 
$^{29}$SiS $J$=6--5 & 106922.982        & 0.63$\times$0.58                      &       194                             &       1                                                                 &       0.03                                                            & 0.4     \\  
$^{30}$SiS $J$=6--5 & 105059.205        & 0.65$\times$0.59                      &       199                             &       1                                                                 &       0.02                                                            & 1.6     \\  
Si$^{34}$S $J$=6--5 & 105941.505        & 0.64$\times$0.59                      &       13                                      &       1                                                                 &       0.02                                                            & 0.2     \\  
CS $J$=2--1                     &  97980.953            & 0.56$\times$0.49                         &       41                                      &       1                                                                 &       0.4                                                             & 5.9     \\  
$^{13}$CS $J$=2--1  &  92494.271        & 0.72$\times$0.66                      &       8                                       &       1                                                                 &       0.008                                                   & 0.14    \\  
C$^{34}$S $J$=2--1  &  96412.952        & 0.78$\times$0.63                      &       32                                      &       1                                                                 &       0.04                                                            & 0.6     \\  
C$^{33}$S $J$=2--1  &  97172.064        & 0.89$\times$0.82                      &       216                             & 2                                                               &       0.005                                                   & 0.09    \\  
\hline
\hline
\multicolumn{7}{p{\textwidth}}{(Col.\,2) Rest laboratory frequency of the line; 
(Col.\,3) size of the synthetic beam;
(Col.\,4) position angle of the beam;
(Col.\,5) 1$\sigma$ rms value of the flux density;
(Col.\,6) maximum value of the flux density;
(Col.\,7) maximum value of the velocity-integrated flux density over the velocity range spanning between $v_\mathrm{*}-v_\mathrm{\infty}$ and $v_\mathrm{*}+v_\mathrm{\infty}$, 
where $v_\mathrm{*}$ is the systemic velocity of the source in the LSR and $v_\mathrm{\infty}$ is the terminal expansion velocity of the CSE (see Section\,\ref{sec:intro}).}\\
\end{tabular}
\end{center}
\end{table*}

\section{Results} \label{sec:res}
In this section we present the observational results derived from the analysis of the brightness distribution and the moment 0 maps for the detected lines of the main isotopologues of SiO, SiS, and CS, 
which are shown in Figures\,\ref{fig:sio_2-1_maps_high}--\ref{fig:cs_2-1_maps}, and their corresponding position-velocity (PV) diagrams shown in Figure\,\ref{fig:main_pvs}.
We present spectra of the four emission lines towards the position of the star extracted from the central pixel of the merged (ALMA plus 30\,m OTF) observations in Figure\,\ref{fig:spec_cpix}. 
Azimuthal averages of the brightness distribution of the ALMA-30\,m merged data of the channel corresponding to the systemic velocity of \irc\ are presented in Figure\,\ref{fig:azave}.
We also present the OTF spectra of the four emission lines towards the position of the star in Figure\,\ref{fig:spec_otf}.
Maps of the brightness distribution and PV diagrams corresponding to the rest of the detected isotopologues are presented in Appendices\,\ref{sec:app_iso_maps} and \ref{sec:app_iso_pvs}.

\subsection{SiO} \label{sec:obsres_sio}
We detected emission from the $J$=2-1 lines of SiO, $^{29}$SiO, and $^{30}$SiO.
The spatial distribution of these emission lines is compatible with a roughly spherical CSE, which extends up to approximately 11\arcsec or equivalently $\sim$600\,\rstar \citep[\rstar=0\arcsecp019$\pm$0.003,][]{rid88}. 
This CSE is radially expanding, as seen in Figure\,\ref{fig:main_pvs}, with a $v_\mathrm{\infty}$ consistent with previous measurements (14.5\,\kms, see Section\,\ref{sec:intro}). 

\begin{figure*}[hbtp!]
\centering
\includegraphics[scale=0.63]{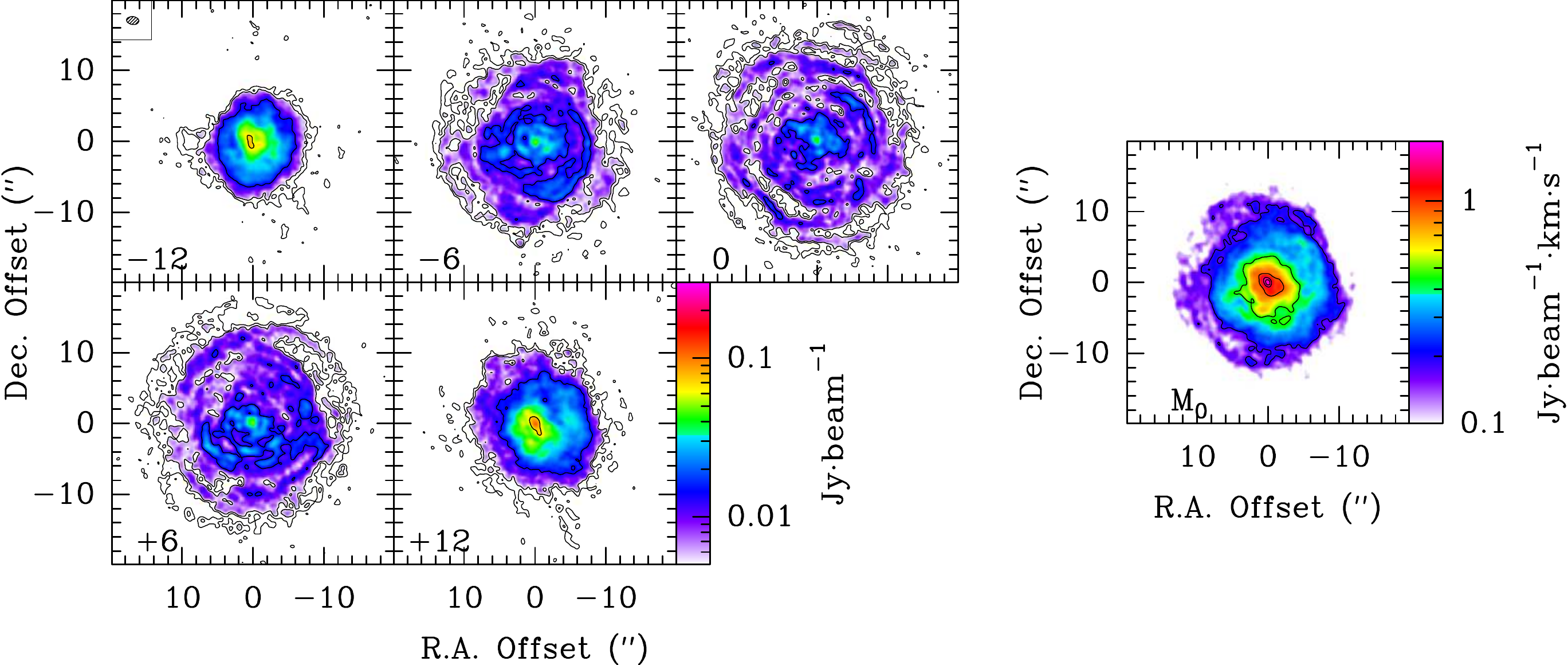}
\caption{SiO $J$=2--1 maps extracted from high spatial-resolution data cube. 
\textit{Left:} Flux density ($S_{\mathrm{\nu}}$) maps at different offset velocities with respect to the systemic velocity of the source \citep[$v_\mathrm{*}\sim$-26.5\,\kms,][]{cer00} in LSR scale.
The central velocity offset of each channel is shown at the bottom-left corner of each panel in kilometres per second.
The width of each velocity channel is approximately 1\,\kms.
The coordinates are given as offsets from the source position in arcseconds (see Section\,\ref{sec:obs}).
The size and orientation of the synthetic beam are shown at the top-left corner inside the first panel.
The contours shown in black correspond to 5$\sigma$, 2\%, 10\%, 50\%, and 90\%\ of the peak flux density (see Table\,\ref{tab:summary}).  
\textit{Right:}  
Moment zero map. 
The contours shown correspond to 10\%, 25\%, 50\%, 75\%, and 90\%\ of the peak emission (see Table\,\ref{tab:summary}).}
\label{fig:sio_2-1_maps_high}
\end{figure*}

SiO emission displays an inhomogeneus spatial distribution, with arcs and clumps (see Figure\,\ref{fig:sio_2-1_maps_high}).
The bulk of the emission arising from the most compact and inner shells, that is $r\lesssim$5\arcsec, seems to be elongated along the north-east-south-west (NE-SW) direction \citep{fon14},
with the brightness distribution peak towards the position of the continuum peak.
From the position of the star, the brightness starts to decrease outwards along the radial direction, 
(as can be seen in the brightness distribution of the emission at the channel corresponding to $v-v_\mathrm{*}$=0\,\kms)
decreasing by a factor of two at $r_\mathrm{2}\sim$0\arcsecp86 ($\sim$45\,\rstar) and by a factor of $e$ at $r_\mathrm{e}\sim$3\arcsec\ ($\sim$160\,\rstar).  
Irregular enhancements of the emission along the radial direction are clearly seen in Figure\,\ref{fig:azave}.
The most noticeable enhancements or bumps appear at approximately
1\arcsecp8 ($\sim$90\,\rstar) and 7\arcsecp8 ($\sim$410\,\rstar), 
and some structures evoke the aspect of circular arms, such as the two structures seen at $r\sim$8\arcsec\ between PA$\sim$[90--180$^\circ$] and
PA$\sim$[270--360$^\circ$], and the two at $r\sim$10\arcsec\ between PA$\sim$[135--180$^\circ$] and PA$\sim$[315--360$^\circ$] (see Figure\,\ref{fig:sio_2-1_maps_high}).

Self-absorption and/or absorption of the continuum emission arising from the inner layers of the CSE is observed at negative velocities for the gas located in front of the star.
This can be noticed in the different brightness temperatures at the most extreme blue and redshifted velocities of the line spectrum towards the position of the star in Figure\,\ref{fig:spec_cpix}.
From this spectrum, we also observe that emission at the horns, and in particular the redshifted horn, is more prominent than emission at the line center. 
This difference, a priori, could be a consequence of the higher opacity of the gas expanding at the terminal velocity. 

\subsection{SiS} \label{sec:obsres_sis}
We detected emission of the $J$=5--4 and 6--5 lines of SiS, $^{29}$SiS, $^{30}$SiS, and Si$^{34}$S.
The spatial distribution of emission of these lines is similar to that of SiO.

\begin{figure*}[hbtp!]
\centering
\includegraphics[scale=0.63]{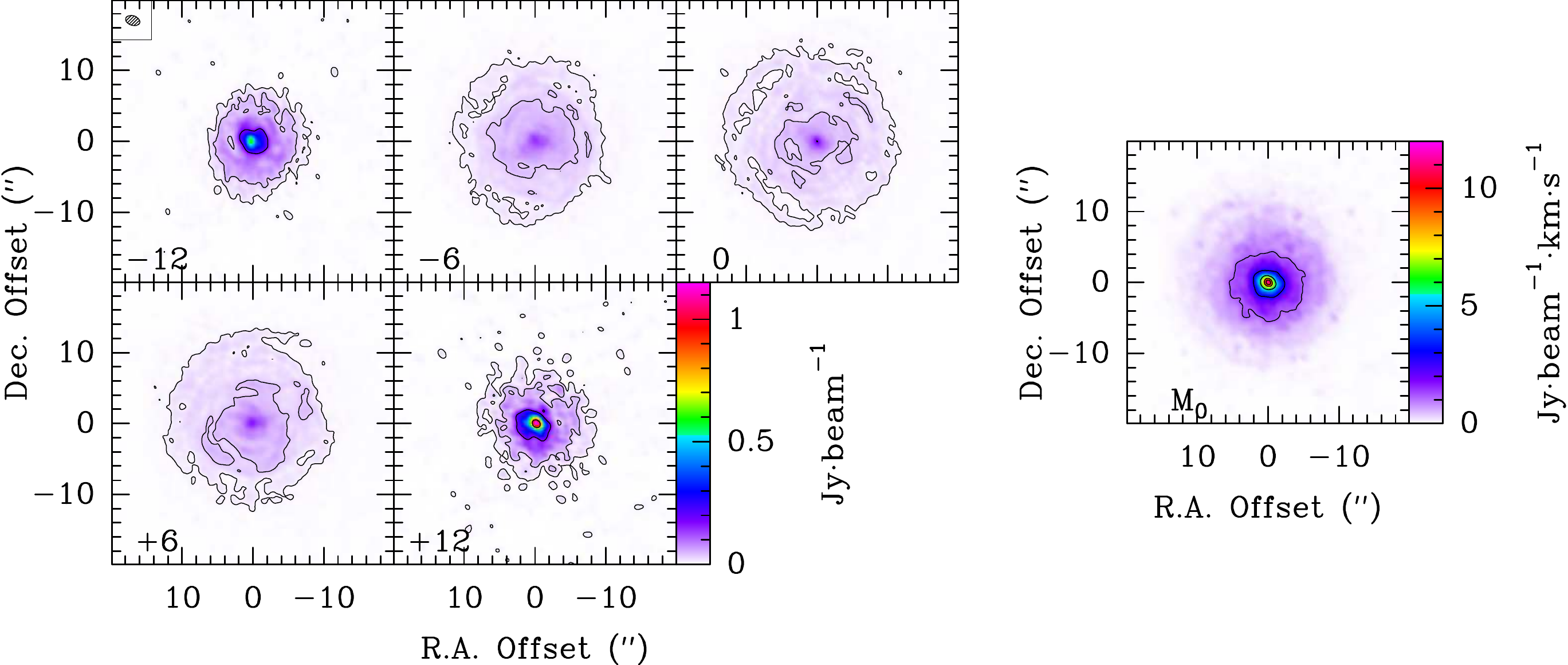}
\caption{As in Figure\,\ref{fig:sio_2-1_maps_high}, but for SiS $J$=5--4.}
\label{fig:sis_5-4_maps}
\end{figure*}

\begin{figure*}[hbtp!]
\centering
\includegraphics[scale=0.63]{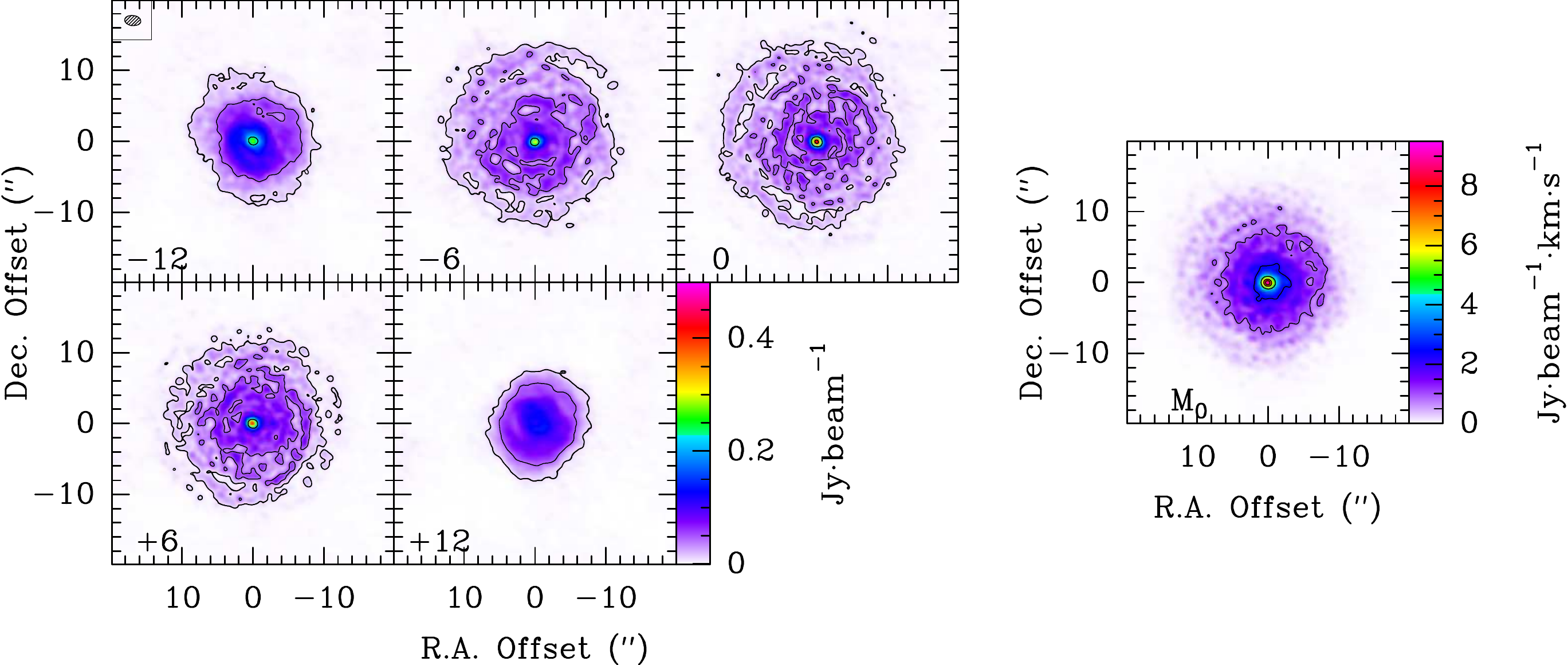}
\caption{As in Figure\,\ref{fig:sio_2-1_maps_high}, but for SiS $J$=6--5.}
\label{fig:sis_6-5_maps}
\end{figure*}

SiS emission is also characterised by an inhomogeneus spatial distribution, displaying clumpy structures and arcs (see Figures\,\ref{fig:sis_5-4_maps} and \ref{fig:sis_6-5_maps}). 
The bulk of the emission arising from the most compact and inner shells ($r\lesssim$5\arcsec) is essentially spherical.
The emission peak at the systemic velocity coincides with the position of the continuum peak.
The brightness distribution observed in this velocity channel displays spatial inhomogeneities that are more clearly seen for the $J$=6--5 emission line (Figure\,\ref{fig:sis_6-5_maps}).
As for SiO, the flux density falls outwards along the radial direction from the position of the star (see Figure\,\ref{fig:azave}).
However, the azimuthal average of the flux density of the SiS $J$=5--4 line displays a decrease of a factor of two at 
$r_\mathrm{2}\sim$1\arcsecp1 ($\sim$60\,\rstar) and of a factor $e$ at $r_\mathrm{e}\sim$1\arcsecp7 ($\sim$90\,\rstar),
while that of the $J$=6--5 line shows $r_\mathrm{2}\sim$0\arcsecp8 ($\sim$40\,\rstar) and $r_\mathrm{e}\sim$0\arcsecp9 ($\sim$50\,\rstar).
The analysis of the azimuthal average of the flux density of both SiS lines leads to the identification of several density enhancements located at 
approximately 4\arcsecp3 ($\sim$225\,\rstar), 5\arcsecp9 ($\sim$310\,\rstar), and
8\arcsecp1 ($\sim$425\,\rstar) for the SiS $J$=5--4 line,
and at 2\arcsecp8 ($\sim$150\,\rstar), 4\arcsecp7 ($\sim$245\,\rstar), and 7\arcsecp9 ($\sim$420\,\rstar)
in the case of the $J$=6--5 line. 
The brightness distribution of the $J$=6--5 emission line at the channel corresponding to the systemic velocity of the source also shows the presence of bright arcs appearing at 
$r\sim$8\arcsec\ between PA$\sim$[270--360$^\circ$] and also at $r\sim$5\arcsec\ between PA$\sim$[20--180$^\circ$] (see Figure\,\ref{fig:sis_6-5_maps}).

The SiS line profiles shown in Figure\,\ref{fig:spec_cpix} display strong differences between them. 
Emission at velocities close to the systemic velocity of the source is similar for the two lines, $J$=6--5 and $J$=5--4.
However, the intense horns seen at terminal velocities in the SiS $J$=5--4 line profile are not visible in the $J$=6--5 line profile. 
This anti-correlation between the line wing emission of the two lines was previously reported by \cite{car90}. 
The redshifted peak of the SiS $J$=5--4 line is approximately ten times stronger than its counterpart in the $J$=6--5 line.
The difference in the intensity of the two horns of the SiS $J$=5--4, and also between the blue and redshifted wings of the $J$=6--5 line, 
may be interpreted as self-absorption and/or absorption of the continuum emission emitted by the inner layers of the CSE, 
as was the case for SiO (see Section\,\ref{sec:obsres_sio} and Figure\,\ref{fig:spec_cpix}).

\subsection{CS} \label{sec:obsres_cs}
\begin{figure*}[hbtp!]
\centering
\includegraphics[scale=0.63]{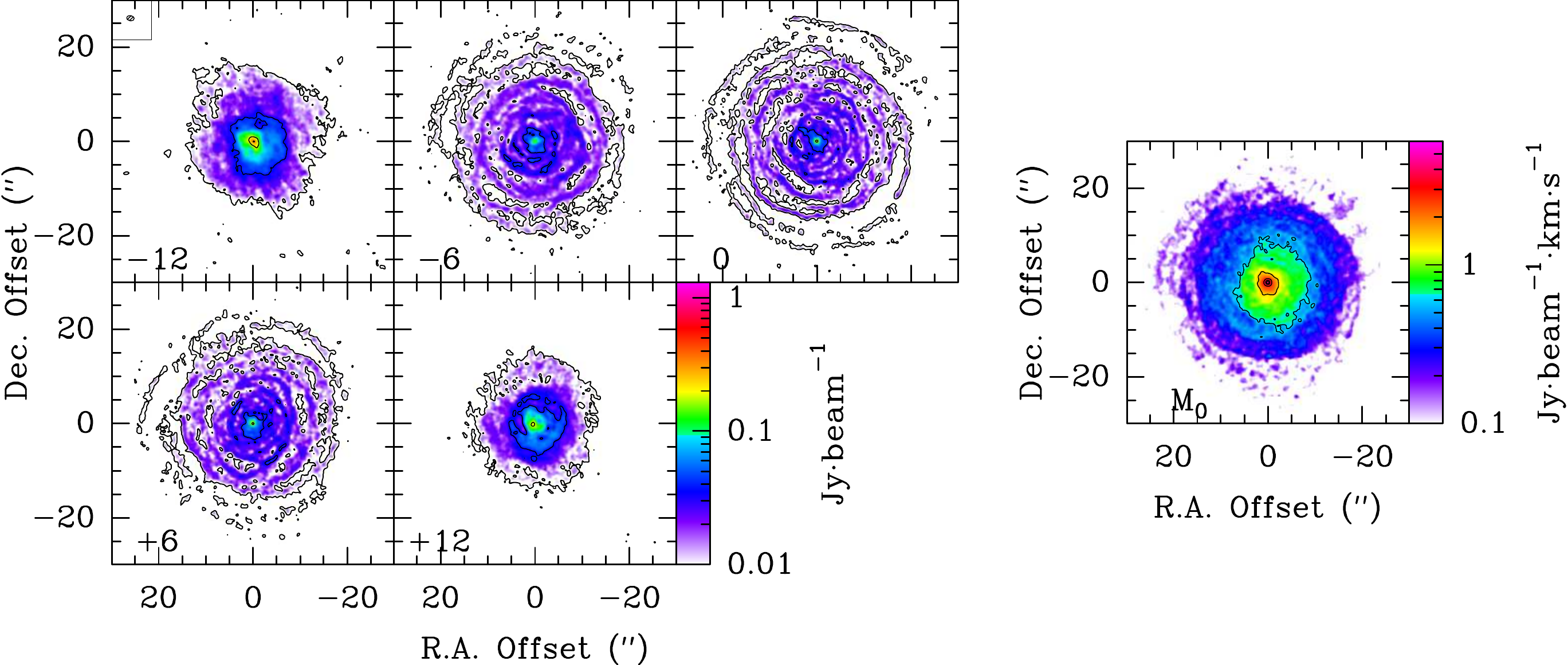}
\caption{As in Figure\,\ref{fig:sio_2-1_maps_high}, but for CS $J$=2--1.
We note that there is a change of the spatial scale compared to that of Figures\,\ref{fig:sio_2-1_maps_high}--\ref{fig:sis_6-5_maps}.}
\label{fig:cs_2-1_maps}
\end{figure*}

\begin{figure*}[hbtp!]
\centering
\includegraphics[scale=0.50]{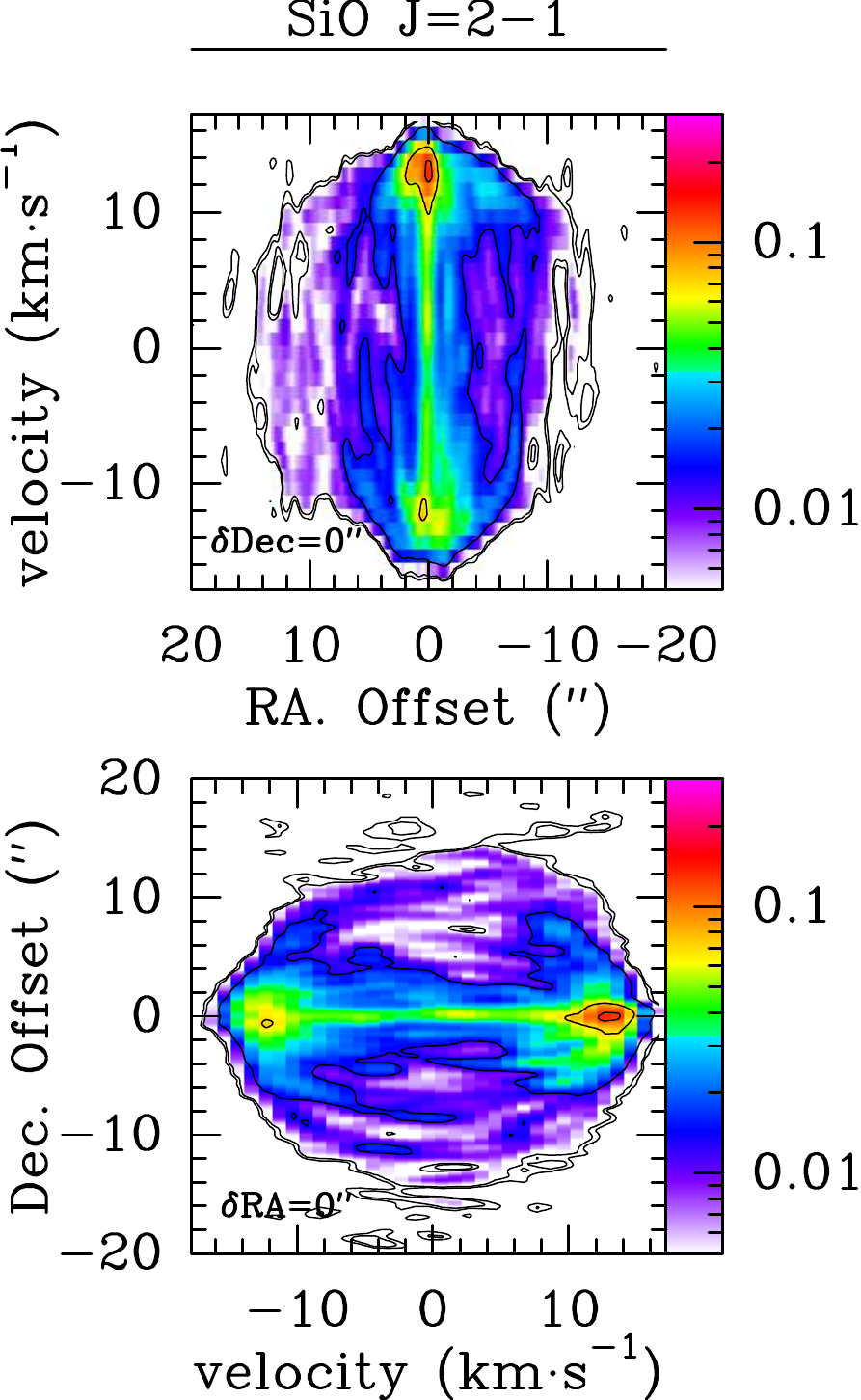}
\includegraphics[scale=0.50]{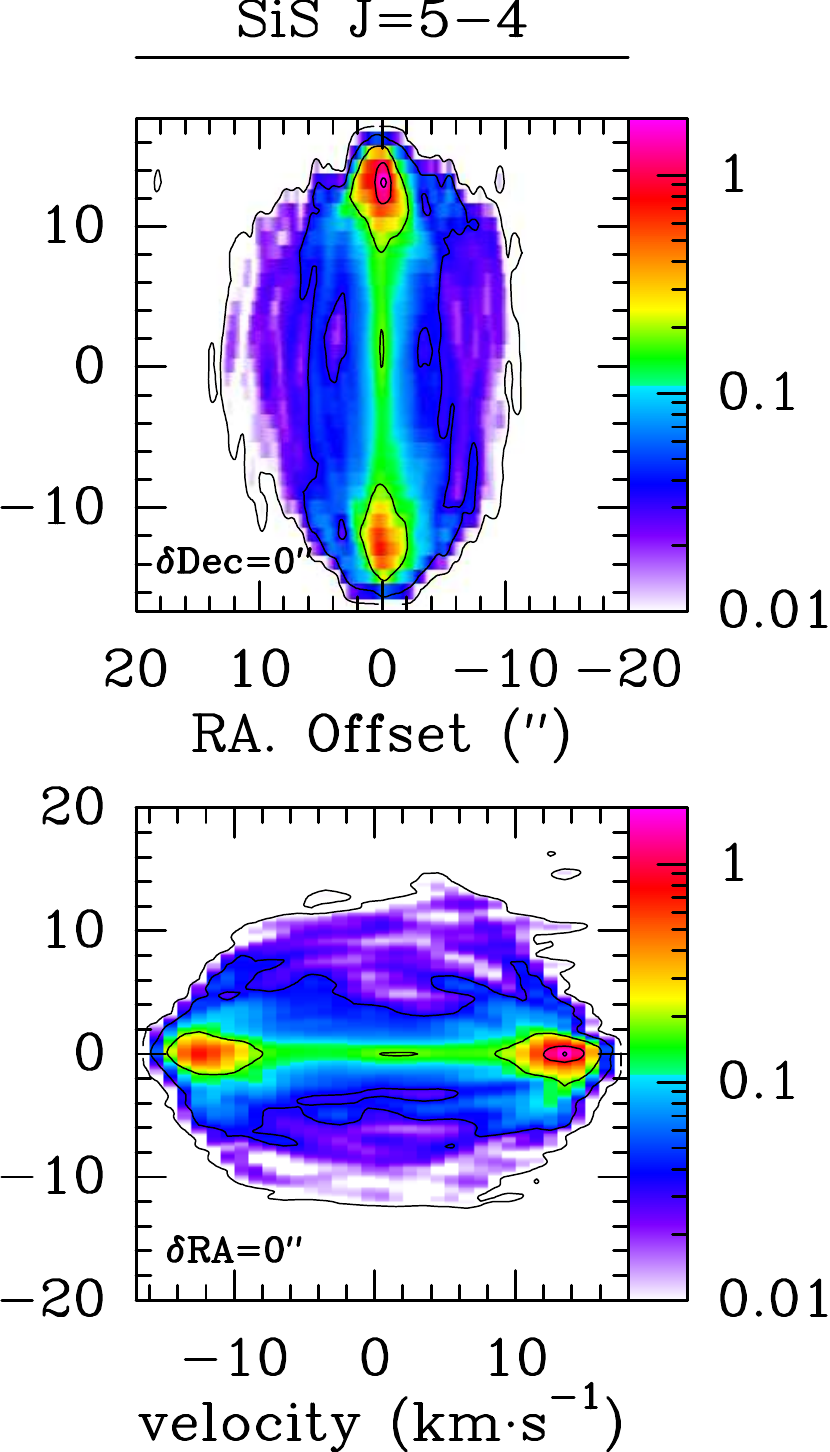}
\includegraphics[scale=0.50]{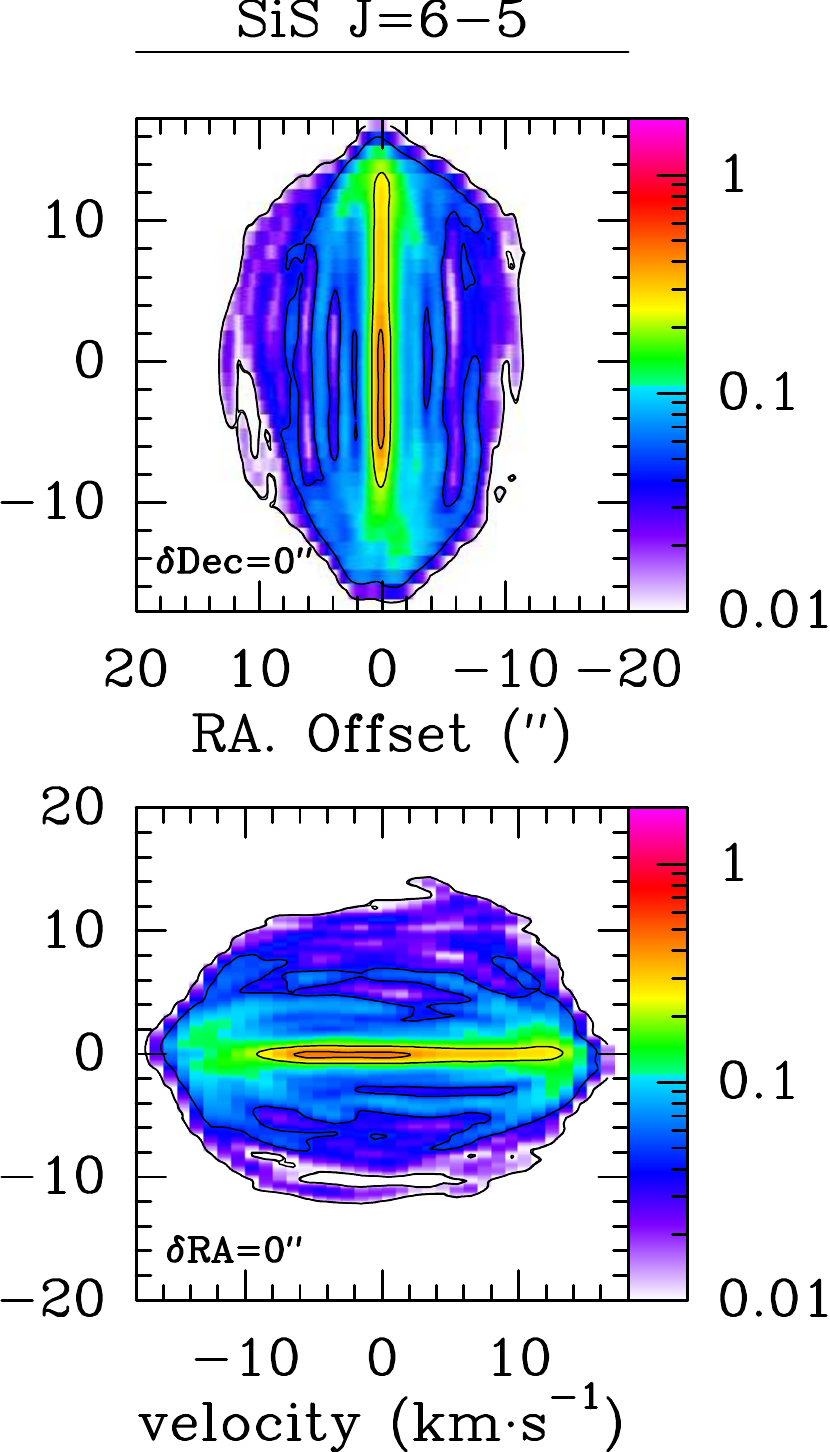}
\includegraphics[scale=0.50]{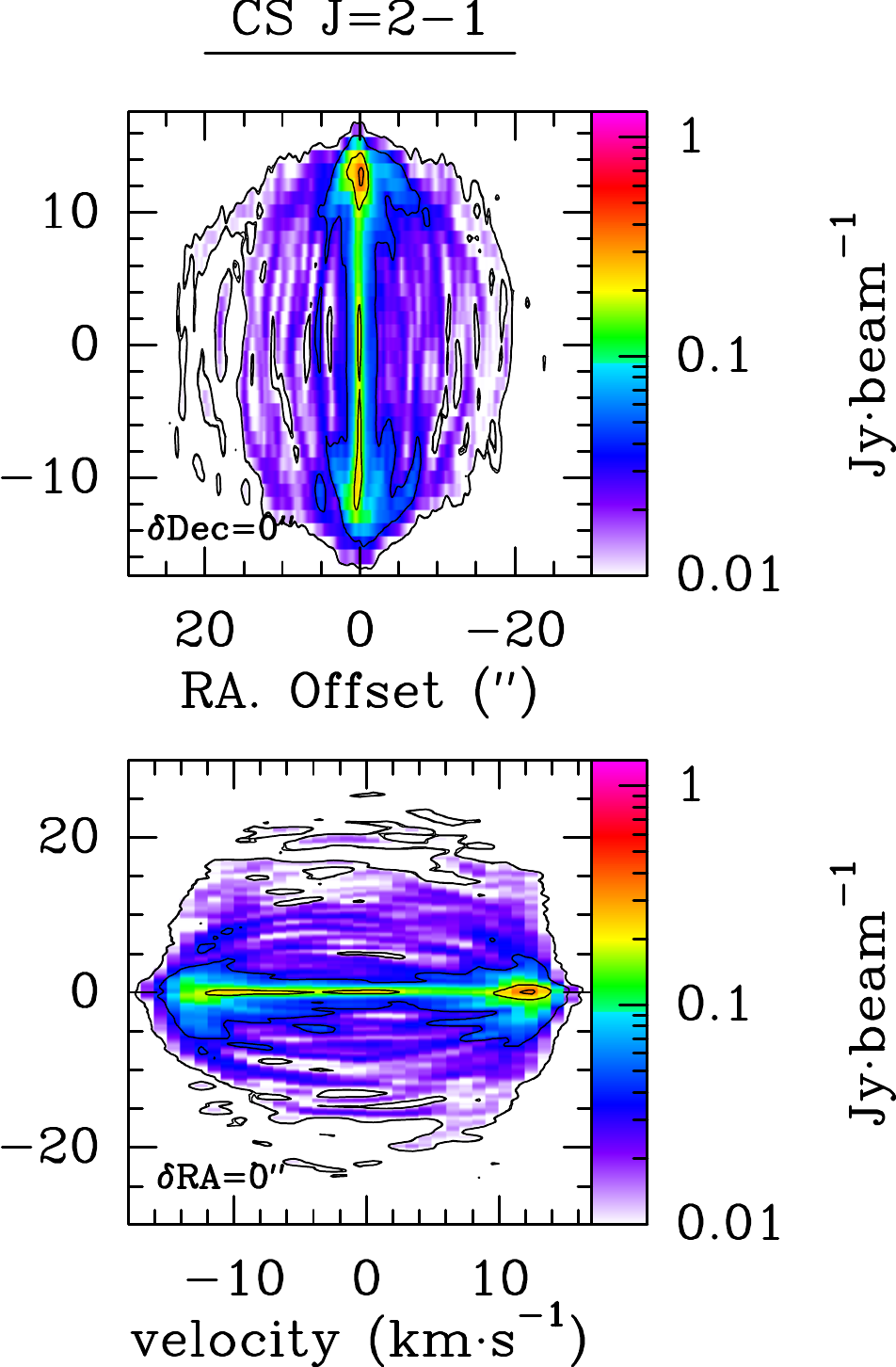}
\caption{Position-velocity (PV) diagrams of detected lines of SiO, SiS, and CS main isotopologues. 
The contours shown in black correspond to 5$\sigma$, 2\%, 10\%, 50\%, and 90\%\ of the peak emission (see Table\,\ref{tab:summary} for details.).
\textit{Top:} 
PV diagram of the flux density is shown corresponding to a plane with a declination offset (see Section\,\ref{sec:obs}) equal to zero.
\textit{Bottom:} 
PV diagram of the flux density is shown corresponding to a plane with a right ascension offset (see Section\,\ref{sec:obs}) equal to zero.
}
\label{fig:main_pvs}
\end{figure*}

We detect emission of the $J$=2--1 lines of CS, $^{13}$CS, C$^{34}$S, and C$^{33}$S.
The emission of the main isotopologue (Figures\,\ref{fig:cs_2-1_maps} and \ref{fig:main_pvs}) 
is compatible with a radially expanding spherical CSE, which extends up to approximately 20\arcsec\ ($\sim$1000\,\rstar) from the star, 
that is much more extended than SiO and SiS emission. 
From the PV diagram the $v_\mathrm{\infty}$ is compatible with previous estimates for the source.

CS emission is remarkably clumpy with sub-structures arranged in the form of concentric shells or circular arcs, as can be seen in Figure\,\ref{fig:cs_2-1_maps}. 
The bulk of the emission arising from the innermost shells ($r\lesssim$5\arcsec) is also circular, as is the case for SiS.
The emission peak at the systemic velocity is on the continuum source.
For this velocity channel, we observe a high degree of clumpiness for distances $r>$10\arcsec\ ($\gtrsim$520\,\rstar) with several thin shells clearly evident in the azimuthal average of the 
flux density (see Figure\,\ref{fig:azave}).
In particular, seven density enhancements were identified through the analysis of the brightness distribution at the systemic velocity, peaking at 
2\arcsecp9 (150$\sim$\,\rstar), 4\arcsecp7 (250$\sim$\,\rstar), 6\arcsecp0 (315$\sim$\,\rstar), 8\arcsecp2 (430$\sim$\,\rstar),
10\arcsecp7 (565$\sim$\,\rstar), 14\arcsecp2 (750$\sim$\,\rstar), and 15\arcsecp8 (830$\sim$\,\rstar) from the central star.
The radius where the flux density has decreased by a factor of two is $r_\mathrm{2}\sim$0\arcsecp6 ($\sim$30\,\rstar) and of a factor of $e$ at $r_\mathrm{e}\sim$0\arcsecp8 ($\sim$40\,\rstar).

\begin{figure}[hbtp!]
\centering
\includegraphics[scale=0.63]{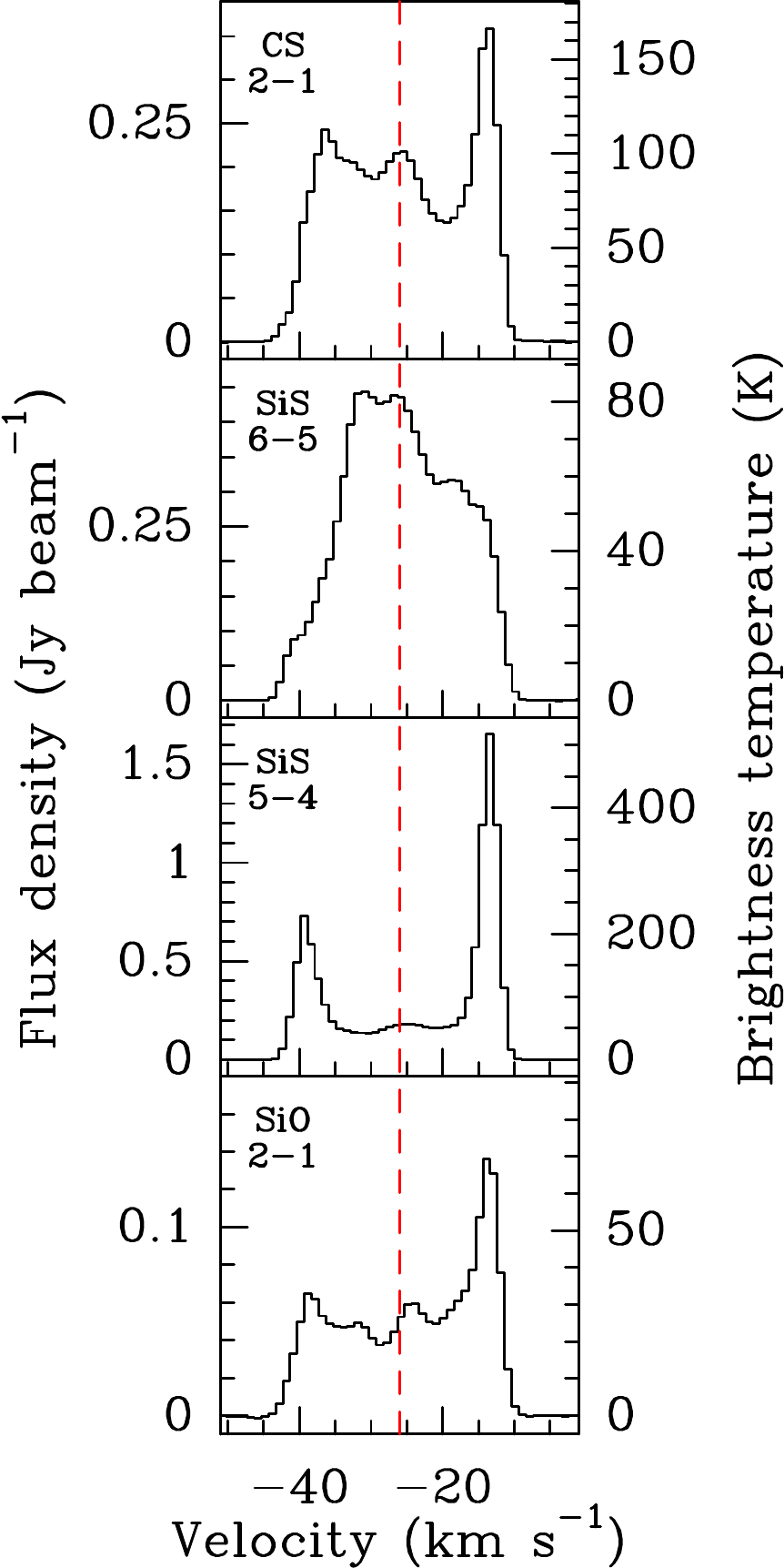}
\caption{Spectrum of the four lines towards the central pixel (0\arcsecp1$\times$0\arcsecp1). This coincides with the position of the star (see Section\,\ref{sec:obs})
obtained from the ALMA-OTF merged observations (see also Table\,\ref{tab:summary} for the basic parameters of the lines).
The dashed red line represents the systemic velocity of the source \citep[$v_\mathrm{*}\sim$-26.5\,\kms,][]{cer00} relative to the LSR.
}
\label{fig:spec_cpix}
\end{figure}

From the spectrum towards the stellar position in Figure\,\ref{fig:spec_cpix}, we observe, as for SiO, that emission at the horns is more prominent than emission at the line center. 
Particularly, the most extreme redshifted emission is $\gtrsim$1.5 times more intense than emission at any other velocity component, possibly as a
consequence of the higher opacity of the gas expanding at the terminal velocity.
As for SiO and SiS, CS-blueshifted emission presents an emission deficit, which could be due to self-absorption and/or absorption of the continuum emission emitted by the inner layers of the CSE
(see Section\,\ref{sec:obsres_sio} and Figure\,\ref{fig:spec_cpix}).

\section{Radiative transfer modelling} \label{sec:mod}
In this section we present the method and results for the radiative transfer modelling of the SiO $J$=2--1, SiS $J$=5--4 and 6--5, and CS $J$=2--1 emission.
We aim to determine the radial abundance distribution of these molecules in the CSE of \irc. 

We carried out radiative transfer models using the code that is described in detail in \cite{agu12}.
The code computes the radiation field and the level populations of a certain molecule by solving the statistical equilibrium equations coupled to the radiation transfer problem following an iterative classical process
under the large velocity gradient (LVG) approximation \citep[][]{sob60,cas70,gol74}.
The emergent line profiles are calculated by ray-tracing across the envelope.

The CSE is modelled as a spherical circumstellar gas and dust envelope expanding at a constant velocity.
It is divided according to a spherical shell gridding of 100 shells evenly spaced on a logarithmic scale for the radius. The physical model of the CSE is essentially an updated revision of the models in \cite{agu12} and \cite{cer13}. 
The updated parameters are the kinetic temperature radial profile and the mass-loss rate, which have been recently estimated in \cite{gue18}.
The fixed parameters of the model are presented in Table\,\ref{tab:modparam}.
The code uses as input the radial abundance profiles that are presented in Figure\,\ref{fig:inpabun}.

The collisional rates taken into account to solve the statistical equilibrium equations consider only collisions with H$_2$. 
Mass correction was applied when necessary. The rates used are the following: 
$i$) for SiO we used the values from \cite{day06} and \cite{bal17}, including ro(tational)-vibrational transitions up to $v$=1, and for the lowest 50 
rotational levels;
$ii$) for SiS we used the values from \cite{klo08} and \cite{tob08} \citep[see also][]{vel15}, including ro-vibrational transitions up to $v$=2, 
and for the lowest 70 rotational levels;
$iii$) for CS we used the values from \cite{den18}, including ro-vibrational transitions up to $v$=1 and for the lowest 30 
rotational levels.
A higher number of vibrational and rotational levels were considered during the modelling but we report only levels that  are up to those values from which significant 
differences do not arise by increasing the number of levels. 
Specific tests of the impact that infrared (IR) pumping could have in our models are described in the following section.

Our purpose is to derive the radial abundance profiles that reproduce the azimuthal averages observed, at least up to a level where model predictions are on average compatible with the 
observations. 
A fine-tuning of the model in order to fit each detail seen in the azimuthal averages would require a significant increase of the gridding points of the model, 
not in line with the spatial resolution achieved (poorer than $\sim$0\arcsecp6 or $\sim$30\,\rstar), 
which would also lead to a significant increase of the computing time. 
Additionally, it can be argued that a perfect fitting could be achieved by modifying the density, the input abundances, and the excitation conditions as a function of radius in different combinations.
Such a degeneracy can lead to misinterpreting the results, moreover, if we consider that chemistry is an additional factor playing a major role.
Therefore, we will give comments on the general aspects seen in the models while keeping in mind this degeneracy, and leaving the physical interpretation for the discussion in Section\,\ref{sec:dis}.

\begin{table*}[hbtp!]
\caption{Stellar and circumstellar parameters of \irc.}
\label{tab:modparam}
\begin{center}
\begin{tabular}{l c c}
\hline\hline
Parameter                                                   & Value                                          & Reference \\
\hline                                                      
Distance ($d$)                                              & 123\,pc                                        & \cite{gro12} \\
Stellar radius ($R_\mathrm{*}$)                             & 4$\times$10$^{13}$\,cm                         & \cite{agu12} \\
Stellar effective temperature ($T_\mathrm{*}$)              & 2330\,K                                        & \cite{rid88} \\
Gas expansion velocity in region I ($v_\mathrm{exp,I}$)     & 5\,\kms                                        & \cite{fon08} \\
Gas expansion velocity in region II ($v_\mathrm{exp,II}$)   & 11\,\kms                                       & \cite{fon08} \\
Gas expansion velocity in region III ($v_\mathrm{exp,III}$) & 14.5\,\kms                                     & \cite{cer00} \\
Microturbulence velocity ($\Delta v_\mathrm{turb}$)         & 1\,\kms                                        & \cite{ski99,deb12} \\
Mass-loss rate (\mloss)                                     & 2.7$\times$10$^{-5}$\,\my                      & \cite{gue18} \\
Gas kinetic temperature ($T_\mathrm{k}$)                    & $T_\mathrm{*}\times (r/R_\mathrm{*})^{-0.68}$  & \cite{gue18} \\
Gas to dust mass ratio ($\rho_\mathrm{g}/\rho_\mathrm{d}$)  & 300                                            & \cite{agu12} \\
Dust condensation radius ($R_\mathrm{c}$)                   & 5\,$R_\mathrm{*}$                              & \cite{agu12} \\
Dust temperature ($T_\mathrm{d}$)                           & 800$\times (r/R_\mathrm{c})^{-0.375}$\,K       & \cite{agu12} \\
\hline                                                      
\hline
\multicolumn{3}{p{\textwidth}}{The expansion velocity is given for three different regions: region I ($r\leq$5\rstar),
region II (5$<r\leq$20\rstar), and region III ($r>$20\rstar).}\\
\end{tabular}
\end{center}
\end{table*}

\begin{figure}[hbtp!]
\centering
\includegraphics[scale=0.63]{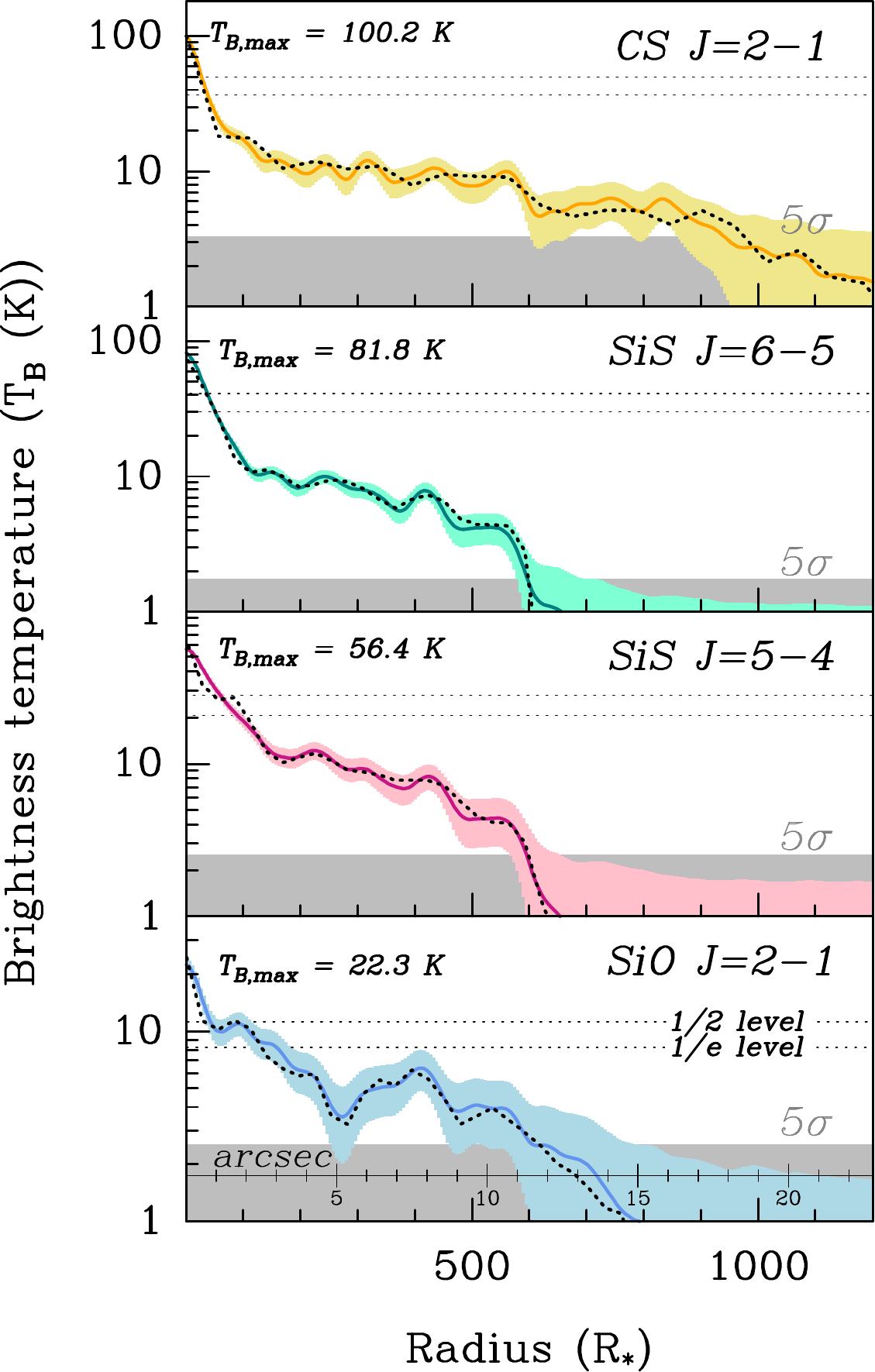}
\caption{Azimuthal average of brightness distribution of four main isotopologue lines detected of CS (yellow line), SiS (green and red lines), and SiO (blue line).
The flux density has been converted to brightness temperature, and the vertical scale is shown in logarithmic scale to improve the visualisation of the plot.
The uncertainty intervals (3$\sigma$) are also shown in each curve with similar colours.  
The spatial scale is shown in units of the stellar radius and we also included an additional axis showing the angular scale in the bottom box.
The equivalence between both scales is 1\,\rstar$\simeq$0\arcsecp019 \citep{rid88}.
On each box, the 5$\sigma$-level detection limit is shown in grey.
We also indicate the maximum value of the brightness temperature for each distribution, which in all cases corresponds to the central pixel. 
We marked the levels in which the temperature decreased by half and by a factor of $e$ of its maximum value.
The results from the radiative transfer models (see Section\,\ref{sec:mod}) are shown as dashed black lines in each box. 
}
\label{fig:azave}
\end{figure}

\begin{figure}[hbtp!]
\centering
\includegraphics[scale=0.63]{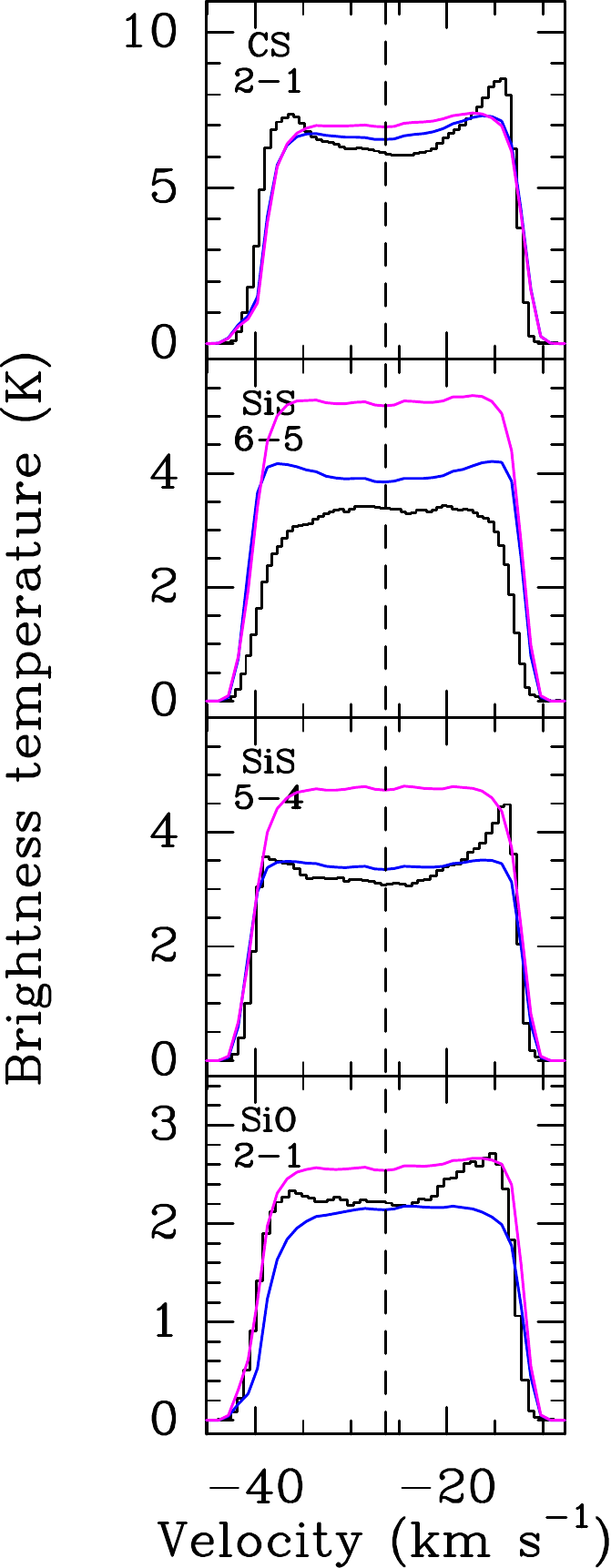}
\caption{Spectrum of  four lines (black histogram) towards star position as observed with IRAM-30\,m telescope. The HPBW for the different frequencies is 
20\arcsecp3 for SiO $J$=2--1,
27\arcsecp1 for SiS $J$=5--4,
22\arcsecp6 for SiS $J$=6--5, and
25\arcsecp1 for CS $J$=2--1.
The coloured curves represent different synthetic spectra created with our radiative transfer code by using different input radial abundance profiles (see Section\,\ref{sec:mod}).
The blue curves represent the model with which we  used the radial abundance profiles shown in Figure\,\ref{fig:inpabun} (solid lines in that figure) as input,
that is, the best fit profiles to the azimuthal average of the brightness distribution of the emission at the systemic velocity of the source from the ALMA and OTF-30\,m merged data.
The curves in magenta represent the model with which we used the same input abundance profile but without including any source of continuum radiation in the model, in order 
to evaluate the impact that IR radiative pumping, produced by the star and the dust, may have in the models.
}
\label{fig:spec_otf}
\end{figure}

\begin{figure}[hbtp!]
\centering
\includegraphics[scale=0.30]{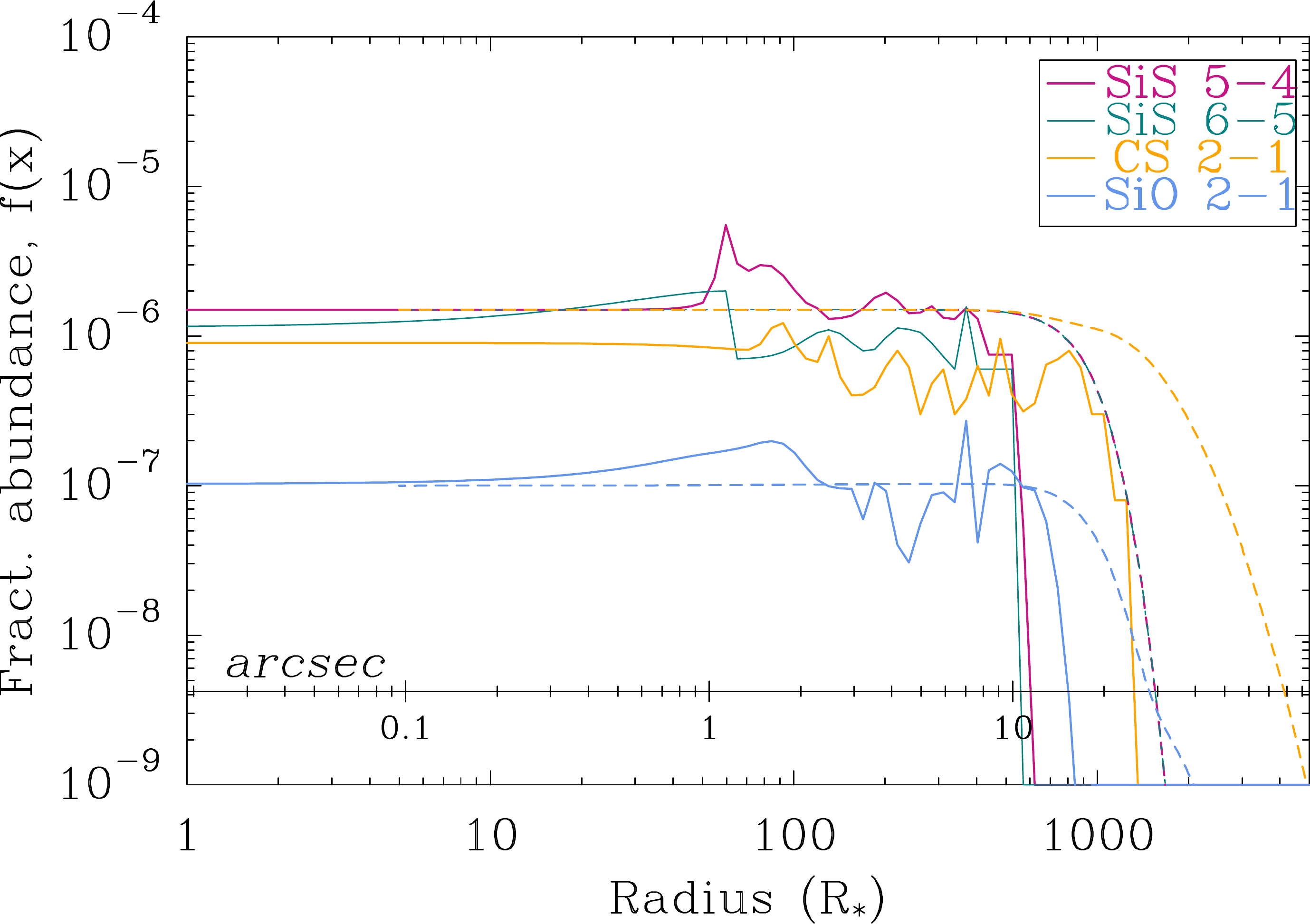}
\caption{Radial abundance profiles (solid lines) used as input to model azimuthal averages in Figure\,\ref{fig:azave}.
The spatial scale is shown in units of the stellar radius and we also included an additional axis showing the angular scale.
The equivalence between both scales is 1\,\rstar$\simeq$0\arcsecp019$\pm$0\arcsecp003 \citep{rid88}.
The different radial profiles are created by using Gaussian components but due to conversion from logarithmic radial sampling the profiles display some abrupt changes.
We note that we have used two different input abundance profiles for SiS, one for each of the lines ($J$=5--4 and $J$=6--5),
aimed at obtaining two independent fits.
Both profiles are very similar, and the true and single radial distribution of SiS may be between both profiles.
The results from the chemical models described in Section\,\ref{sec:chem} are represented by dashed lines of the same colours, 
except for the SiS line where both magenta and green colours are used.
The fractional abundance value is relative to H$_2$.
}
\label{fig:inpabun}
\end{figure}

The results from the radiative transfer modelling can be seen in Figure\,\ref{fig:azave}, where they are compared with the azimuthal averages, 
and in Figure\,\ref{fig:spec_otf}, where they are compared with the spectra towards the position of the star as observed with the IRAM-30\,m telescope. 
The half-power beam width (HPBW) of the 30\,m telescope as a function of the frequency ($\nu$) can be calculated as: 
HPBW{[\arcsec]}$\sim$2460/$\nu${[GHz]\footnote{\tt http://www.iram.es/IRAMES/mainWiki/Iram30mEfficiencies}.}

The fit to the azimuthal averages of the brightness distribution of the channel corresponding to the systemic velocity of the source is reasonably well achieved 
by using the radial abundance profiles presented in Figure\,\ref{fig:inpabun}.
For a radially expanding envelope at constant velocity, the emission in this channel is representative of the gas approximately distributed over the plane of the sky, 
and whose movement is contained within this plane since its velocity component parallel to the line of sight is equal to zero when referenced 
to the systemic velocity of the source.

From the observations and the radiative transfer models, the first molecule to disappear from the gas phase is SiS (at $\sim$600\,\rstar), 
then SiO (at $\sim$700--800\,\rstar), and significantly farther, CS (at $\sim$1000\,\rstar).
These values are consistent with those reported in previous studies, where the authors reported sharp cutoffs in the radial distribution 
at $\sim$500--750\,\rstar\ (for SiS and SiO) and $\sim$900--1000\,\rstar\  \citep{bie93,agu12}.
The fractional abundances of SiO and CS seem to have a mild decrease in the intermediate CSE (between $\sim$100 and 300\,\rstar) compared to their abundances 
in the innermost CSE. 
For SiO, we derive a fractional abundance of $\sim$10$^{-7}$, which is in good agreement with the values reported by \cite{agu12}. 
Several studies of SiO emission have reported higher fractional abundances in the innermost CSE of \irc\ complemented by the analysis of IR ro-vibrational lines \citep{kea93,sch06}.
However, \cite{fon14} reported a lower abundance of an order of magnitude ($\sim$10$^{-8}$) very close to the star ($r<$5\,\rstar) based on the analysis of the SiO $J$=6--5 emission as observed with 
high-angular resolution ($\sim$0\arcsecp25) with the Combined Array for Research in Millimetre-wave Astronomy (CARMA).
In the case of CS, we derive an average fractional abundance of $\sim$10$^{-6}$ in the innermost CSE, decreasing down to $\sim$5$\times$10$^{-7}$ between 
$\sim$100 and 300\,\rstar, and then keeping this value in the outermost CSE. 
The abundance profile derived in the current work is in good agreement with that derived by \cite{agu12}, 
who also detected an abundance decline in the intermediate envelope. 
However, our abundance close to the star is a factor of two higher than theirs.
We note that these authors considered the first 50 rotational levels within the vibrational states $v$=0--3, while we have considered only the first 30 rotational levels in $v$=0--1. 
For SiS, we find an average abundance of $\sim$1--2$\times$10$^{-6}$ in the region up to $\sim$70--100\,\rstar, mildly decreasing from that point by less than a factor of two up to 
the cutoff radius ($\sim$600\,\rstar).  
Previous studies of SiS emission towards \irc\ reported radial abundance distributions compatible with our results \citep{bie93,agu12,fon15}.

In order to reproduce some of the emission enhancement shells seen in the azimuthal average of the four lines, the fractional abundance has to be increased locally at certain radii from the star. 
As we have noted before, this can also be achieved by increasing the total gas density without a significant increase of the fractional abundance of the given molecule.
At this point, we cannot conclude which of the two scenarios is more realistic until our results are also tested against chemical predictions (Section\,\ref{sec:chem}).
We also want to note that our study is based solely on the analysis of just a single line in the cases of SiO and CS, and two lines in the case of SiS, which prevents us from 
drawing more solid conclusions on the physical conditions of the CSE and the spatial distribution of the three species. 
Dedicated subarsecond resolution observations of different rotational lines, probing different excitation conditions, 
would be required to derive a more realistic model.

Following the fitting procedure of the radial brightness distribution of the line emission representing the gas expanding in the plane of the sky, 
we tested the accuracy of the model to predict the observed spectrum of the emission lines towards \irc\ as observed with the IRAM-30\,m antenna.
The results obtained from this model are presented in Figure\,\ref{fig:spec_otf}.
In this case, the gas producing the spectrum corresponds to the gas distributed along the line of sight within the telescope beam. 
For a radially expanding spherical CSE with the velocity gradient given in Table\,\ref{tab:modparam}, the gas
 directly behind the star should be moving away from us, thus, it is seen in the 
red part of the spectrum, while the gas in front of the star should be moving towards us, and is therefore seen in the blue part of the spectrum.
The synthetic spectra of the SiO $J$=2--1, SiS $J$=5--4, and CS $J$=2--1 reproduce (on average) the intensity of the observed lines at central velocities ($v<$\vinf) using the same radial abundance profile as 
the one we used for the azimuthal averages. 
In the case of the SiS $J$=6--5 line, the synthetic spectrum shows on average a difference of $\sim$20\% in the intensity compared to the observed line at central velocities ($v<$\vinf).
At terminal velocities, the differences between the observed and synthetic line profiles are more noticeable.

There is a clear difference between the amount of gas probed by the azimuthal average and the amount of gas probed by the 30\,m spectrum. 
In the latter case, the column density of the gas is much higher, which could explain the differences observed between the two cases.
Moreover, since all the gas located between the dust condensation region ($\sim$20\,\rstar) and the outermost parts of the CSE is expanding at the terminal expansion velocity along the line of sight, 
most of this gas emission will lie either at the most extreme red or blue velocities in the spectra towards the position of the star. 
This means that in this case the spatial resolution is significantly hampered,
which also explains up to a certain degree the discrepancies seen in the horns of the lines.
Likewise, \cite{bie93} reported difficulties when simultaneously fitting the SiS $J$=6--5 radial average profile and the line-of-sight spectrum convolved to the beam of a single-dish telescope with a single abundance profile.
Nevertheless, we want to recall that our model is, on average and within uncertainties, satisfactory.

We tested the impact that IR pumping has on the computed intensities in our model. 
Through this mechanism, the population of rotational levels may deviate from what they would have in the absence of the IR radiation emitted by the star and the dust.
Moreover, considering the variable nature of AGB stellar emission, this effect is coupled to the stellar phase, although, this may have a limited impact on optically thick, low $J$  lines \citep{cer14}.
In our model we eliminated all sources of continuum radiation, that is the star and the dust, to test this scenario.
In the case of the azimuthal average of the brightness distribution of the ALMA-30\,m channel map at the velocity equal to the systemic velocity of the source, 
moderate effects are observed. 
In the case of SiO, the predicted emission is overestimated by $\lesssim$15\%\ in the region between $\sim$100 and 300\,\rstar with respect to the model including sources of IR radiation. 
In the outermost layers ($\gtrsim$400\,\rstar) the predicted emission is understimated by $\lesssim$10\%.
For CS, the predicted emission of the non-IR scenario is overestimated by $\lesssim$10\%\ between $\sim$100 and 300\,\rstar. 
For SiS, the predicted emission would be $\lesssim$20\%\ overestimated from $\sim$100\,\rstar\ up to the cutoff radius.

We also compared the synthetic spectra obtained from our non-IR model with the observed spectra towards \irc\ with the IRAM-30\,m telescope (see Figure\,\ref{fig:spec_otf}).
 In terms of average intensity,
predicted emission for CS is very similar to that of the scenario where sources of IR radiation are included. 
However, SiO and SiS synthetic lines produced with the non-IR model are overestimated by a factor of $\sim$1.3.

According to our calculations, the excitation conditions for the four transitions are dominated by 
collisions in the innermost region of the CSE ($\lesssim$10\,\rstar).
From that point, the excitation temperatures increase above the kinetic temperature and then change to sub-thermal for 
SiO (at $\sim$250\,\rstar), CS (at $\sim$300\,\rstar), and SiS (at $\sim$600\,\rstar).
In the case that IR pumping is not considered, the four transitions become sub-thermal at approximately the same distances except for the SiS 
transitions, which change to sub-thermal at much closer distances to the star ($\sim$200\,\rstar).
In terms of opacities, our models predict moderate to high opacities throughout the envelope for the four lines analysed.

The case of the excitation of the two SiS transitions, $J$=5--4 and 6--5, is particularly interesting, moreover considering the significant differences between the two line profiles in Figure\,\ref{fig:spec_cpix}. 
Our model does not include line overlaps from other abundant species at IR wavelengths, and we limited the calculations to the lowest 70 rotational levels in the vibrational levels $v$=0, 1, and 2.
Therefore, we cannot assert to what extent the IR line overlaps,  nor if the variability of IR stellar light contributes to the excitation of the SiS lines analysed here. 
However, the line profiles shown in Figure\,\ref{fig:spec_cpix} suggest that the $J$=5 level is involved in a particular excitation mechanism that makes $J$=5--4 emission at terminal velocities considerably 
intense, while it is suppressed at the same velocities for the $J$=6--5 line.
The most plausible explanation for this scenario is maser emission of the $J$=5--4 line, probably as a consequence of an IR line overlap as we discuss below.

SiS excitation has been investigated by other authors in the context of the observational study of SiS emission towards \irc\ \citep[][and references therein]{olo82,sah84,car90,bie93,fon06,fon18}.
\cite{olo82} reported observations of the SiS $J$=5--4 line with the Onsala-20\,m telescope (HPBW$\sim$42\arcsec), showing a double-peaked U-shaped profile similar to that seen in Figure\,\ref{fig:spec_otf}.
\cite{sah84} observed SiS $J$=6--5 emission with the Number Two 10.4\,m telescope of the Owens Valley Radio Observatory (OVRO) with a HPBW of $\sim$64\arcsec\, and the 
$J$=5--4 emission with the 14\,m telescope of the Five College Radio Astronomy Observatory (FCRAO) with a HPBW of $\sim$58\arcsec.
From the analysis of the data, \cite{sah84} found that a population inversion occurred for the $J$=5--4 transition in the inner regions of the CSE, due to IR pumping. 
\cite{sah84} argued that the horns shown in the double-peaked $J$=5--4 profile reported by \cite{olo82}, contrary to the rounded profiles that \cite{sah84} reported for the rest of 
the SiS lines they analysed, was a consequence of weak maser emission observed at maximum stellar phase with the smallest beam among all the observations reported so far at that time.
\cite{sah84} also investigated the impact of IR line overlaps on the excitation of SiS that could overexcite the $J$=5--4 line, 
in particular due to the overlap of HCN and C$_2$H$_2$ with SiS ro-vibrational lines. 
This line overlapping scenario was also discussed by \cite{bie93}, but they found no obvious lines of HCN or C$_2$H$_2$ overlapping SiS lines that could have a significant impact on the population 
of low $J$ SiS levels.
\cite{car90} carried out time monitoring of the SiS $J$=4--3, 5--4, and 6--5 emission towards \irc\ with the 20\,m telescope in Onsala. 
They reported that there was correlation between the IR flux and the line shape of the $J$=5--4 and 6--5 lines, 
with the relative intensity of the $J$=6--5 line decreasing at stellar maximum, while that of the $J$=5--4 line increased at the same stellar phase.
However, they did not find correlation for the $J$=4--3 line. 
This would indicate that the population of the $J$=5 level varies in phase with the IR light curve.
\cite{fon06} reported the first detection of high $J$ maser emission and they proposed that this is produced due to line-overlapping between SiS ro-vibrational lines and mid-IR lines of C$_2$H$_2$ and HCN.
\cite{fon18} showed that the SiS $J$=5--4 line displays narrow peaks at high expansion velocities, which vary strongly with the pulsation phase, while the $J$=6--5 line shows no intense peaks 
with a nearly constant profile over the stellar period.
The occurrence of these varying peaks in the $J$=5--4 line and their lack in the $J$=6--5 line leads these authors to conclude that the peaks are masers in nature. 

\section{Chemistry} \label{sec:chem}
In this section we present the chemical model we created aimed at testing the observational and radiative transfer results.
The model considers a spherical CSE expanding isotropically in the radial direction at constant velocity (equal to \vinf) from an initial radius equal to 2$\times$10$^{14}$\,cm.
Several molecules are used as input for the model, in particular, the parent species listed in Table\,\ref{tab:parent},
which are abundant species formed in the innermost layers of the CSE that will determine the subsequent chemistry. 
The abundances taken for these species are presented in Table\,\ref{tab:parent}, which are the values used in \cite{agu17} after updating the abundances of CS, SiO, and SiS according to our results 
from the radiative transfer analysis.
The chemical network encompasses about 8\,000 gas-phase, photo-induced, and cosmic-ray reactions, with rate constants taken from the literature and from the UMIST and KIDA databases \citep{mce13,wak15}.  
We assume that the CSE is externally illuminated by the local UV radiation field \citep{dra78}.
The adopted cosmic-ray ionisation rate of H$_2$ is 1.2$\times$10$^{-17}$\,s$^{-1}$ \citep{agu06}.
In this work we have adopted a $N_{\mathrm{H}}$/$A_{\mathrm{v}}$ value that is 1.5 times lower than the classical value determined for the local ISM 
\citep[1.87$\times$10$^{-21}$\,cm$^{-2}$mag$^{-1}$,][]{boh78}.
This assumption has been made on the basis of reproducing the interferometer observations of the region where photochemistry occurs \citep[][and references therein]{agu17}.
Particularly for the three species analysed in this work, we updated two important reactions in the chemistry of SiS according to \cite{zan18}, which are the destruction of SiS with atomic oxygen 
and the reaction between Si and SO.
For the photo-dissociation rates of the studied species we took the values from \cite{pat18} for CS and from \cite{hea17} for SiO. 
In the case of SiS, there are no estimates of its photo-dissociation rate, thus as an educated guess, we assumed the same value than that of SiO.

\begin{table}[hbtp!]
\caption{Abundances of  parent species in  chemical model.}
\label{tab:parent}
\begin{center}
\begin{tabular}{l c c}
\hline\hline
Molecule & Value & Reference \\
\hline                                                      
CO                      & 6.0$\times$10$^{-4}$ & \citep{agu12} \\ 
C$_2$H$_2$       & 8.0$\times$10$^{-5}$ & \citep{fon08} \\ 
CH$_4$              & 3.5$\times$10$^{-6}$ & \citep{kea93} \\ 
C$_2$H$_4$  & 8.2$\times$10$^{-8}$ & \citep{fon17} \\ 
H$_2$O      & 1.0$\times$10$^{-7}$ & \citep{dec10} \\ 
N$_2$       & 4.0$\times$10$^{-5}$ & \citep{agu17} \\ 
HCN         & 4.0$\times$10$^{-5}$ & \citep{fon08} \\ 
NH$_3$      & 2.0$\times$10$^{-6}$ & \citep{has06} \\ 
CS          & 1.5$\times$10$^{-6}$ & This work     \\  
H$_2$S      & 4.0$\times$10$^{-9}$ & \citep{agu12} \\ 
SiS         & 1.5$\times$10$^{-6}$ & This work     \\ 
SiO         & 1.0$\times$10$^{-7}$ & This work     \\
SiC$_2$     & 2.0$\times$10$^{-7}$ & \citep{cer10} \\ 
SiH$_4$     & 2.2$\times$10$^{-7}$ & \citep{kea93} \\ 
PH$_3$      & 1.0$\times$10$^{-8}$ & \citep{agu14} \\ 
HCP         & 2.5$\times$10$^{-8}$ & \citep{agu07} \\ 
\hline                                                      
\hline
\multicolumn{3}{p{\columnwidth}}{The fractional abundances are relative to H$_2$.}\\
\end{tabular}
\end{center}
\end{table}

The predictions from the chemical model are presented in Figure\,\ref{fig:inpabun}.
The most extended distribution is that of CS, while SiO and SiS abundances fall at similar distances from the star given that 
we use the same photo-dissociation rate for both molecules.
These predictions are qualitatively similar to the observations and the results from the radiative transfer analysis (see Figures\,\ref{fig:azave} and \ref{fig:inpabun}), 
in terms of the order of the abundance fall-off of the three species.
There are relatively small discrepancies in the values of the distance where the abundances of these molecules start to decrease as predicted by the chemical models. 
In particular, these predictions are on average below a factor of 1.5, being the fall-off distances of the chemical models larger than the observational measurements.
As we discuss below, these discrepancies emerge mainly from the simplicity of the chemical model, 
while the calibration uncertainties and the choice of a robust weighting factor of 0.5 in the imaging process (Section\,\ref{sec:obs}) have only a limited impact. 

The chemical model fails in reproducing the observed high-density shells, with the three species essentially keeping frozen-out abundances until they are photo-dissociated at the outer CSE.
It can be argued that the simplicity of the model is probably the main factor responsible for the inconsistencies found between the observations and the chemical models.
First, the interaction between the gas and the dust particles is not considered, which has an important impact in the depletion of the gas from the dust condensation region outwards,
as well as in other chemical processes throughout the whole CSE. 
There is also an important source of uncertainty in the choice of the UV field and the $N_{\mathrm{H}}$/$A_{\mathrm{v}}$ factor.
This ratio has been fixed to be a factor of 1.5 lower than the value by \cite{boh78}, in order to reproduce the results obtained for other molecules such as the carbon chains presented 
in \cite{agu17}.
Higher values of the UV field illuminating the CSE or lower values of the $N_{\mathrm{H}}$/$A_{\mathrm{v}}$ ratio than those considered in the models 
would shift the photo-dissociation radius of the three species to inner regions of the CSE.
As we discuss in Section\,\ref{sec:dis}, this latter consideration could also describe the evolution of mass loss from the star. 

There is also another important source of uncertainty, which is the adopted value for the photo-dissociation rate of SiS, which was assumed to be equal to that of SiO. 
In the case of the abundance fall-off distance of SiO compared to that of CS, the observational results are consistent with the adopted values for the 
photo-dissociation rates 
\citep[1.6$\times$10$^{-9}$\,exp(-2.66\,$A_{\mathrm{v}}$) and 3.7$\times$10$^{-10}$\,exp(-2.32\,$A_{\mathrm{v}}$) in s$^{-1}$ for SiO and CS, respectively;][]{hea17,pat18}
predicting that SiO should photo-dissociate closer to the star than CS. 
If the observed fall-off distances respond to a pure photo-dissociation process, then the observations may support that the photo-dissociation rate of SiS should be a bit higher than 
that of SiO. 
This observational constraint should be considered in future research aiming to derive accurate SiS photo-dissociation rates.

\section{Discussion} \label{sec:dis}
The brightness distribution of the different lines we have analysed display important characteristics, such as the arcs and shells seen or the differences in their extension. 
For this difference we provide an interpretation based primarily on our radiative transfer and chemical modelling results. 

\subsection{Arcs and shells: episodic and isotropic mass loss?}
As noted in Section\,\ref{sec:obs}, the spatial distribution of the four lines here studied present deviations from how an homogeneous spherical wind should appear.
Different arcs and shells are seen in the brightness distributions shown in Figures\,\ref{fig:sio_2-1_maps_high}--\ref{fig:cs_2-1_maps}, 
which could be a consequence of different physico-chemical factors. 
The azimuthal average of the four brightness distributions at the systemic velocity channel (see Figure\,\ref{fig:azave}) also display several brightness enhanced shells.
Some of these shells appear at almost equal distances for the four lines studied, in particular, 
one is clearly seen at $\sim$8\arcsecp0 ($\sim$420\,\rstar), 
and others are seen in some of the azimuthal averages at $\sim$2\arcsecp8 ($\sim$150\,\rstar), 6\arcsecp0 ($\sim$315\,\rstar), and 10\arcsecp6 ($\sim$560\,\rstar).
These brightness enhanced shells may seem not exactly coincidental when comparing the four lines displaying small radial shifts that could be due to non-strict circularity of the shells and 
also possible excitation or optical thickness effects.
For the shells appearing at $\sim$420\,\rstar, ruling out excitation effects after the radiative transfer analysis, two different interpretations may be given:
it is either a chemical effect or the effect of an enhancement of the gas density in a shell.

The chemistry in this region is mainly driven by gas-phase chemistry, photo-induced chemistry due to UV interstellar radiation field, cosmic-ray interaction, and gas-dust interaction. 
Our models, which do not take into account gas-dust chemistry, do not predict a particular enhancement in this region. 
The lack of a detailed model including surface chemistry in dust grains prevents us from assessing the importance that a chemical effect may have in the formation of a chemical enhancement at $\sim$420\,\rstar.
However, there is no observational evidence for a critical physical process occurring at this distance modifying the subsequent chemistry, 
for example a significant increase of the temperature that may enhance thermal desorption from the dust grains.
Therefore, it is reasonable to suppose that the gas-dust interaction should behave in a similar way in the region where the shells appear than immediately near inter-shell regions.
The second and most likely scenario would be an enhancement of the gas and dust density, maybe as a consequence of an enhanced mass-loss event in the past.
Considering an average expansion velocity of the gas equal to $v_\mathrm{\infty}$ and the distance to the star, 
the kinematical age for a shell located at $\sim$420\,\rstar\ would be $\sim$365\,years.

\begin{figure}[hbtp!]
\centering
\includegraphics[scale=0.38]{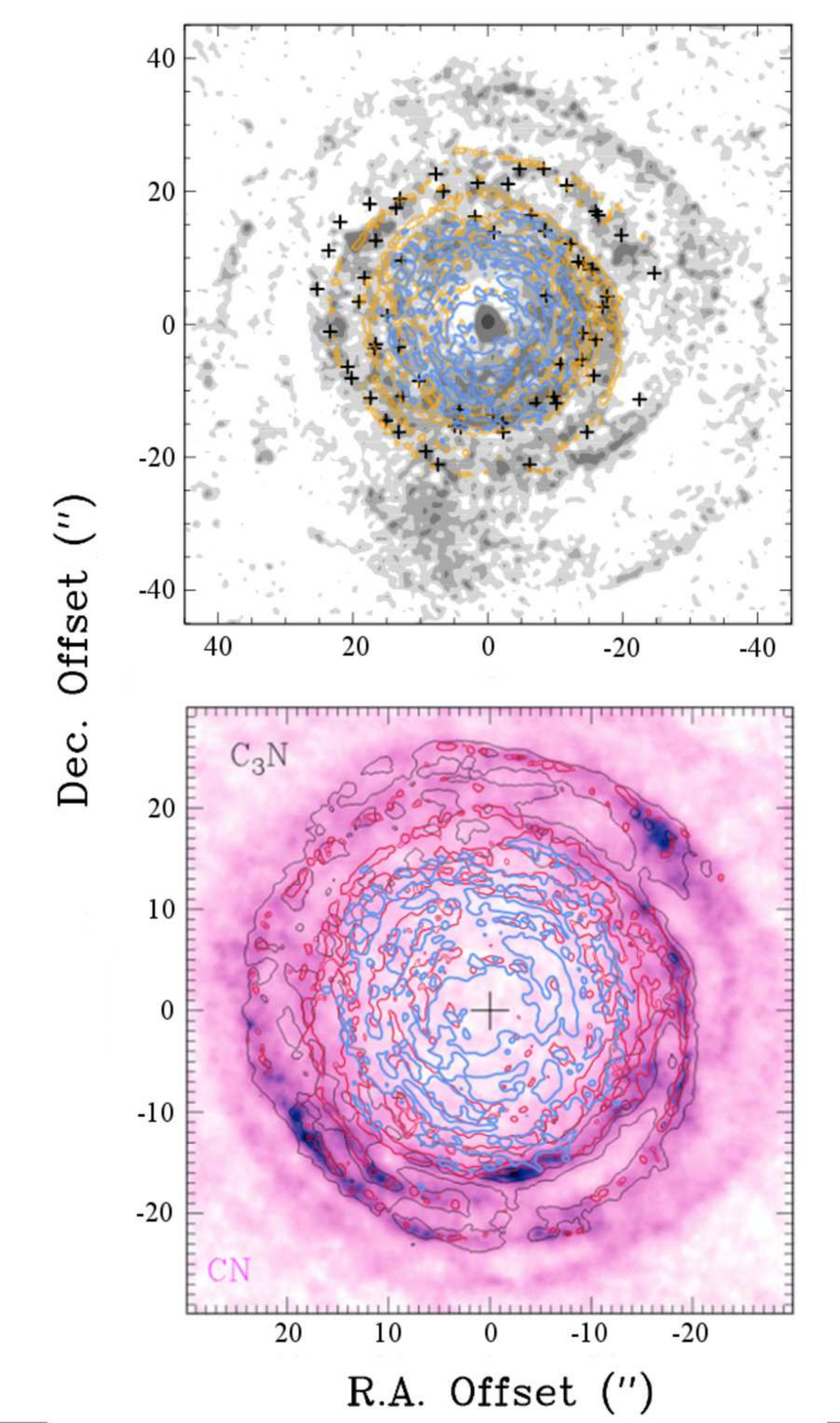}
\caption{\textit{Top}: Comparison between the gas spatial distribution of the $J$=2--1 CS and SiO emission in orange and blue contours, respectively,
and the dust distribution reported in \cite{mau99} as seen in $V$+$B$, where the dark arcs and shells trace the location of the dust 
(background image in black and white colour scale). The small crosses in the image are also from \cite{mau99}, and represent the emission peaks of other molecular 
tracers, namely CN, HNC, and HC$_3$N reported in \cite{luc95} and \cite{luc99}.
\textit{Bottom}: The same contours representing $J$=2--1 CS and SiO emission, with $J$=2--1 CS contours in red (changed to improve the visualisation) 
and $J$=2--1 SiO in blue,
plotted over C$_3$N $N$=10--9 (black contours) and CN $N$=1--0 (background image in pink colour scale) as reported by \cite{agu17}. 
We note that the spatial scale between both figures is different. 
}
\label{fig:comparison}
\end{figure}

Episodic mass loss from central stars in the AGB phase is known to occur for several C-stars \citep{olo96}, including \irc\ \citep{mau99,mau00,cer15}.
\cite{mau99} reported the presence of dust arcs and shells that might be tracing over-dense regions.
These authors also discussed the spatial coincidence of dust and gas shells and arcs occurring at distances of $r\sim$10\arcsec, 15\arcsec\ and 20\arcsec\ (see Figure\,\ref{fig:comparison}),
with a clear double-ring structure to the north-west of the star.
We compared the structures seen in our observations with those traced by dust as seen in the visible scattered light \citep{mau99,lea06} and 
by other molecules such as CO \citep{gue18}, and the species C$_2$H, C$_4$H, C$_6$H, CN, C$_3$N, HC$_3$N, and HC$_5$N \citep{agu17}.

The time lapse between the observations in \cite{mau99} and ours is about 16.5 years, 
which would account for a $\sim$0\arcsecp4 radial shift due to the expansion of the circumstellar material at a constant velocity equal to \vinf.
Ruling out this small shift, the distribution of the dust shells reported by \cite{mau99} is remarkably coincident with that of 
the gas as traced by the lines reported in our work (see Figure\,\ref{fig:comparison}).
Several shell structures seen in CS emission at $r\sim$15\arcsec\ and 20\arcsec\ to the north-west, and at $r\sim$15\arcsec\ and 20\arcsec\ to the east
clearly match with the shells traced by the dust.
The same occurs for SiO emission at distances $r\lesssim$10\arcsec, where several structures of gas and dust are spatially
coincident.
The same spatial correlation is seen for the outermost shells of CS, CN, and C$_3$N emission (Figure\,\ref{fig:comparison}).
These structures were also seen as spatially coherent in more recent observations of CO emission and other carbon-chains listed before \citep{agu17,gue18}.
Therefore, the existence of over-dense shells of gas and dust surrounding \irc\ seems to be the most plausible explanation 
to the brightness enhanced shells seen in the brightness distributions of SiO, SiS and CS. 

Additionally, the presence of a binary companion to \irc\ has been proposed in previous works \citep[][]{gue93,cer15}.
This could result in enhanced episodes of mass loss motivated by the orbital motion of the system when passing through the periastron, 
as discussed by these authors.
The effect of a binary companion would also cause anisotropies and inhomogeneities in the spatial distribution of the dust and the gas.
Other authors have also pointed out the possible importance of the action of different instabilities leading to the 
formation of incomplete dust arcs and clumps \citep{woi06}.  

\subsection{Emission and photo-dissociation size}
The size of the molecular envelope associated with an AGB star depends on the resilience that molecules have against dissociation, 
which in the case of the outermost layers of the CSE is driven by photo-dissociation by UV photons from the ISM.
The results from our work indicate that the observed sizes are slightly smaller compared to the predictions from our chemical model but within uncertainties. 
Beyond all the simplifications made in the model, which have been commented on in Section\,\ref{sec:chem}, there is also a physical explanation 
to the incompatibilities that emerged from the analysis.
The photo-dissociation radius would shift inwards in the case that the envelope is less dense than that of the model and also if it has a clumpy structure, 
or if the UV field is more intense than the value we have adopted (see Section\,\ref{sec:chem}).
To our knowledge, there is no evidence that would suggest that \irc\ is immersed in a region with a UV field stronger than the average UV interstellar radiation field \citep{mau03}.

Following our discussion in the previous section, all the results would be consistent with episodic mass loss from \irc\ together with a clumpy structure.
According to this scenario, UV photons from the ISM could penetrate deeper in the envelope. 
This will activate photo-induced processes at distances closer to the star than those expected with an homogeneous CSE, that is a CSE created with a constant mass-loss rate.
Periods of low mass-loss activity followed by enhanced episodes would explain the inconsistencies described in Section\,\ref{sec:chem},
where we considered a constant mass-loss rate. 
According to the radial abundance profiles presented in Figure\,\ref{fig:inpabun}, brightness enhanced shells are fitted by enhancing the fractional abundance by a factor $\sim$2--3.
Without a complete chemical model that includes gas-dust interaction we cannot affirm to what extent the chemistry contributes to this effect, or even if any chemical contribution is made at all. 

\section{Conclusions} \label{sec:conc}
The observations and analyses we have presented pose strong evidences of a variable mass-loss process acting on \irc\ with timescales of hundreds of years.
In particular, several shells or at least partial shells are seen in the maps obtained with ALMA and the IRAM-30\,m telescope 
(Figures\,\ref{fig:sio_2-1_maps_high}--\ref{fig:cs_2-1_maps})
as well as in the azimuthal average of the emission from the velocity channel equal to the systemic velocity of the object (Figure\,\ref{fig:azave}).
Several episodes of enhanced mass loss may have occurred to create these dense shells. 

We derived the radial profiles of the fractional abundances of SiO, SiS, and CS, which present oscillations that probably reflect the episodic mass-loss of the star.
On average, the fractional abundances estimated are f(SiO)$\sim$10$^{-7}$, f(SiS)$\sim$10$^{-6}$ and f(CS)$\sim$10$^{-6}$.
With these results from the radiative transfer analysis, we have created a standard chemical model based on gas-phase, photo-induced, and cosmic-ray induced processes.
As parent molecules, the three species analysed keep roughly constant abundances within the intermediate CSE until they are photo-dissociated in the more external layers of the envelope.
Our model is able to predict a photo-dissociation radii spatial sequence that is consistent with the observations and the radiative transfer analysis.
This means that SiS is photo-dissociated closer to the star; SiO photo-dissociates slightly farther away but very close to SiO; and CS, which presents the most extended distribution, photo-dissociates much farther away from the star.
However, the chemical model predicts fall-off distances $\sim$1.5 times larger than those derived from the observations and radiative transfer 
analysis, and does not predict oscillations in the chemical abundances of the three species in the intermediate envelope, 
which is probably a consequence of the simplicity of the physical and chemical model.
The local departures from the average fractional abundances have to be considered carefully when making a detailed, spatial-scale analysis,
until a complete chemical model that includes surface chemistry is considered. 

Further research is required in order to understand the processes that lead to the formation of the different shells and clumps observed in the molecular gas of \irc.
From the observational point of view, high-angular resolution observations with ALMA aiming to a resolution equivalent to $\lesssim$5\,\rstar\ would be necessary to resolve the 
innermost regions of the CSE.
From the theoretical side, future efforts should be directed to improve the chemical networks available by adding gas-dust interaction.

\begin{acknowledgements}
We want to thank the referee, J.\,Bieging, for his careful revision of the manuscript which substantially helped improving the quality of this study. 
Our team acknowledges the support given by ERC through the grant ERC-2013-Syg-610256 ``NANOCOSMOS'', the Spanish MINECO through the grants AYA2012-32032 and AYA2016-75066-C2-1-P and 
the CONSOLIDER-Ingenio program ``ASTROMOL'' CSD 2009-00038.
LVP acknowledges support from the Swedish Research Council and the ERC consolidator grant 614264.
This work is based on observations carried out under projects numbers 014-13 and 055-15 with the IRAM-30\,m telescope. IRAM is supported by INSU/CNRS (France), MPG (Germany) and IGN (Spain).
This paper makes use of the following ALMA data: ADS/JAO.ALMA\#2013.1.00432.S.
ALMA is a partnership of ESO (representing its member states), NSF (USA) and NINS (Japan), together with NRC (Canada) and NSC and ASIAA (Taiwan), in cooperation with the Republic of Chile. 
The Joint ALMA Observatory is operated by ESO, AUI/NRAO and NAOJ. 
This research has made use of NASA's Astrophysics Data System. 
This work has made use of GILDAS\footnote{\url{http://www.iram.fr/IRAMFR/GILDAS}} and CASA\footnote{\url{https://casa.nrao.edu/}} softwares to reduce and analyse data.
\end{acknowledgements}

\clearpage
\appendix
\section{Brightness distribution for the rest of the isotopologues}\label{sec:app_iso_maps}
Here we present the brightness distribution and moment zero images for the rest of the detected lines of isotopologues:
the $J$=2--1 line of $^{29}$SiO and $^{30}$SiO; 
the $J$=5--4 and 6--5 lines of $^{29}$SiS, $^{30}$SiS, and Si$^{34}$S;
and the $J$=2--1 line of $^{13}$CS, C$^{34}$S, and C$^{33}$S.

\begin{figure*}[h]
\centering
\includegraphics[scale=0.50]{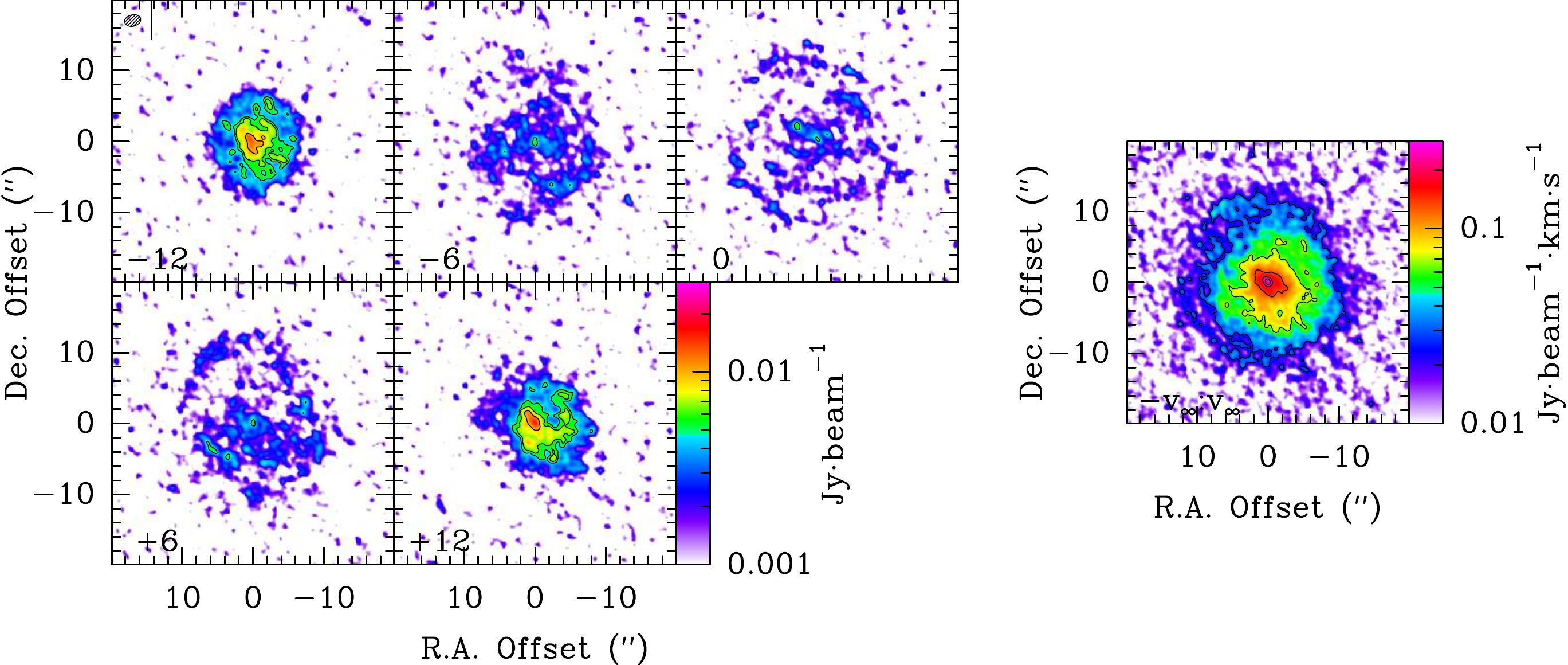}
\caption{$^{29}$SiO $J$=2--1 maps extracted from low spatial-resolution data cube. 
\textit{Left:} Flux density ($S_{\mathrm{\nu}}$) maps at different offset velocities with respect to the systemic velocity of the source \citep[$v_\mathrm{*}\sim$-26.5\,\kms,][]{cer00} in LSR scale.
The central velocity offset of each channel is shown at the bottom-left corner of each panel in kilometres per second.
The width of each velocity channel is approximately 1\,\kms.
The coordinates are given as offsets from the source position in arcseconds (see Section\,\ref{sec:obs}).
The size and orientation of the synthetic beam are shown at the top-left corner inside the first panel.
The contours shown in black correspond to 5$\sigma$, 25\%, 50\%, and 90\%\ of the peak flux density (see Table\,\ref{tab:summary}). 
\textit{Right:} Moment zero map.
The contours shown correspond to 10\%, 25\%, 50\%, 75\%, and 90\%\ of the peak emission (see Table\,\ref{tab:summary}).}
\label{fig:29sio_2-1_maps}
\end{figure*}

\begin{figure*}[h]
\centering
\includegraphics[scale=0.50]{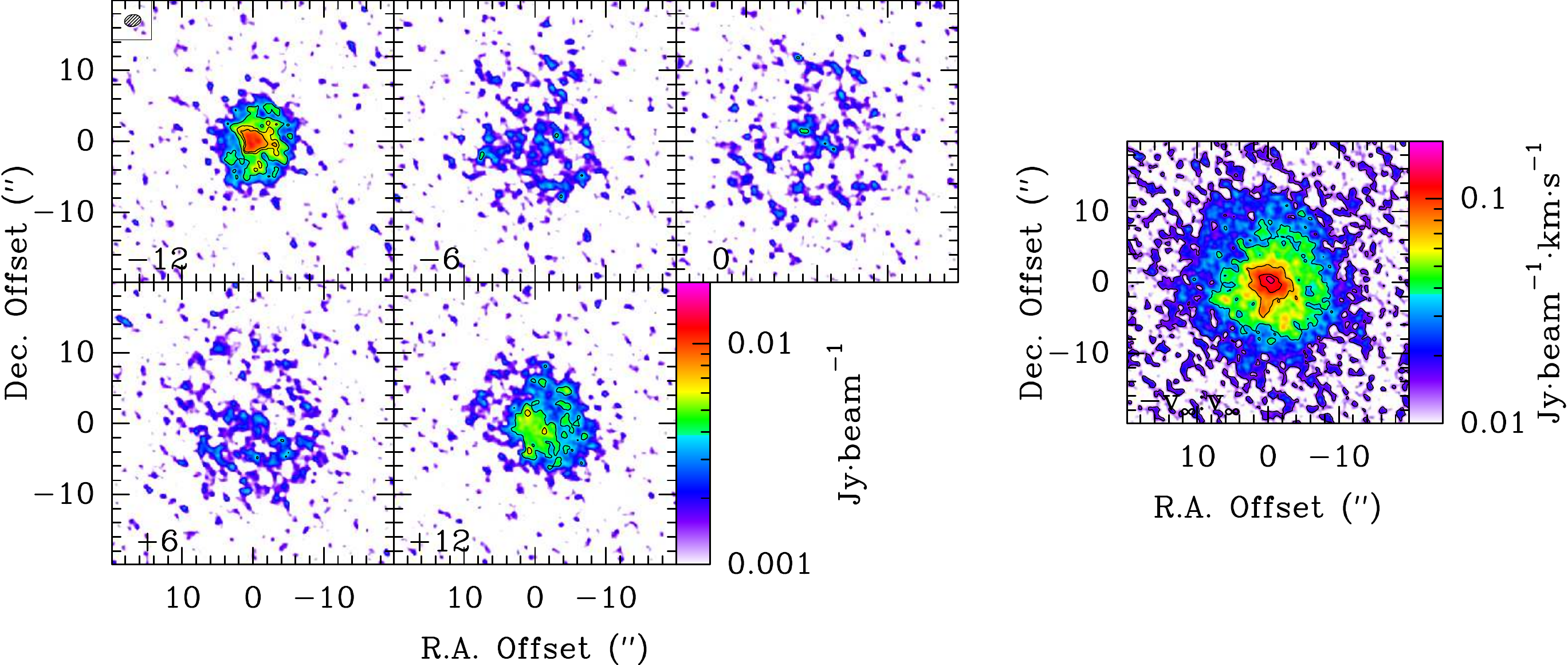}
\caption{As in Figure\,\ref{fig:29sio_2-1_maps}, but for $^{30}$SiO $J$=2--1.}
\label{fig:30sio_2-1_maps}
\end{figure*}

\begin{figure*}[h]
\centering
\includegraphics[scale=0.50]{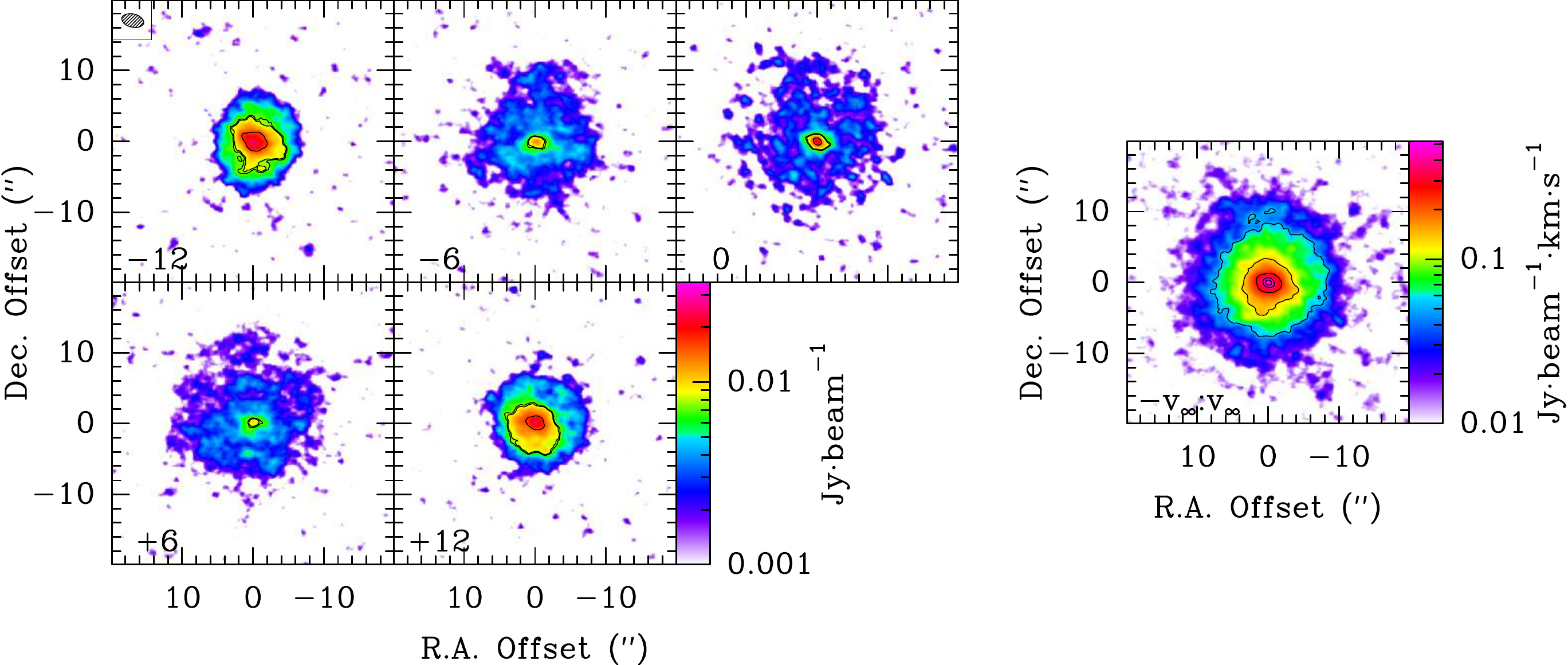}
\caption{As in Figure\,\ref{fig:29sio_2-1_maps}, but for $^{29}$SiS $J$=5--4.}
\label{fig:29sis_5-4_maps}
\end{figure*}

\begin{figure*}[h]
\centering
\includegraphics[scale=0.50]{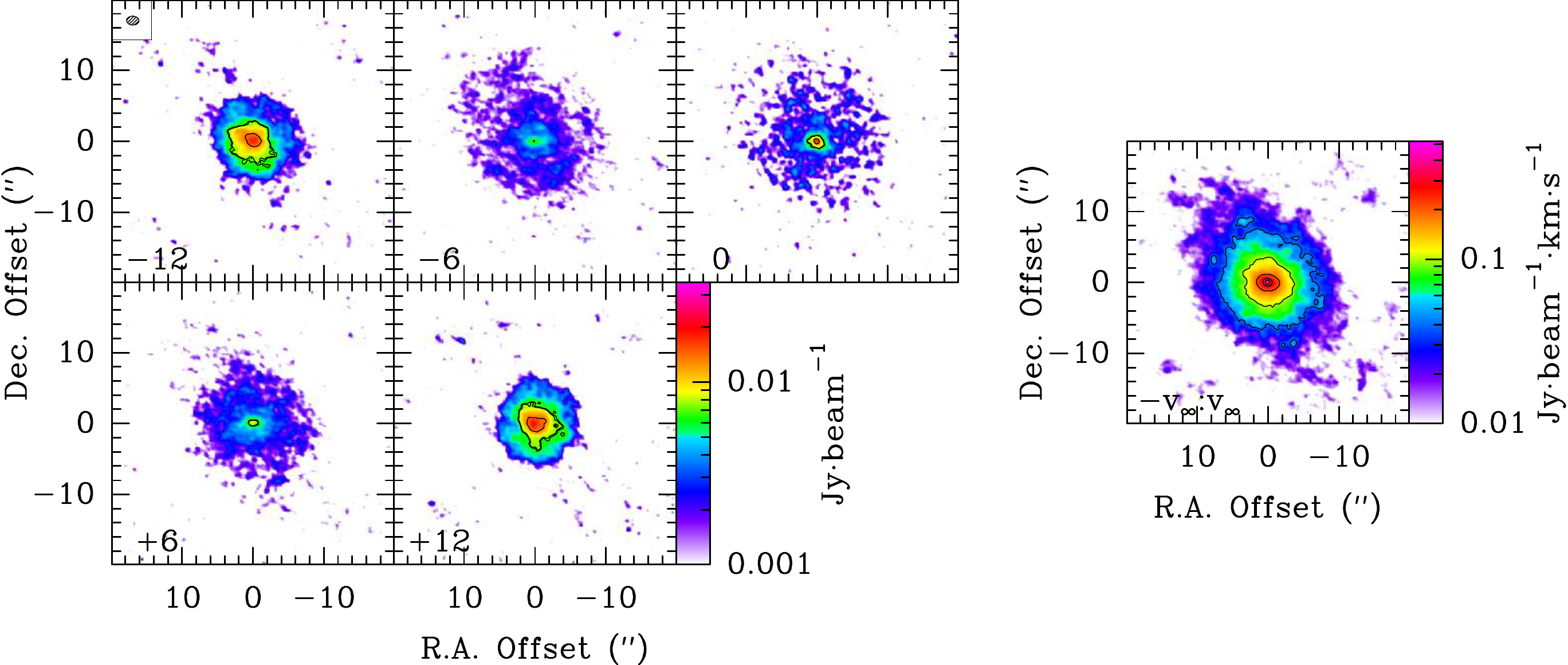}
\caption{As in Figure\,\ref{fig:29sio_2-1_maps}, but for $^{29}$SiS $J$=6--5.}
\label{fig:29sis_6-5_maps}
\end{figure*}

\begin{figure*}[h]
\centering
\includegraphics[scale=0.50]{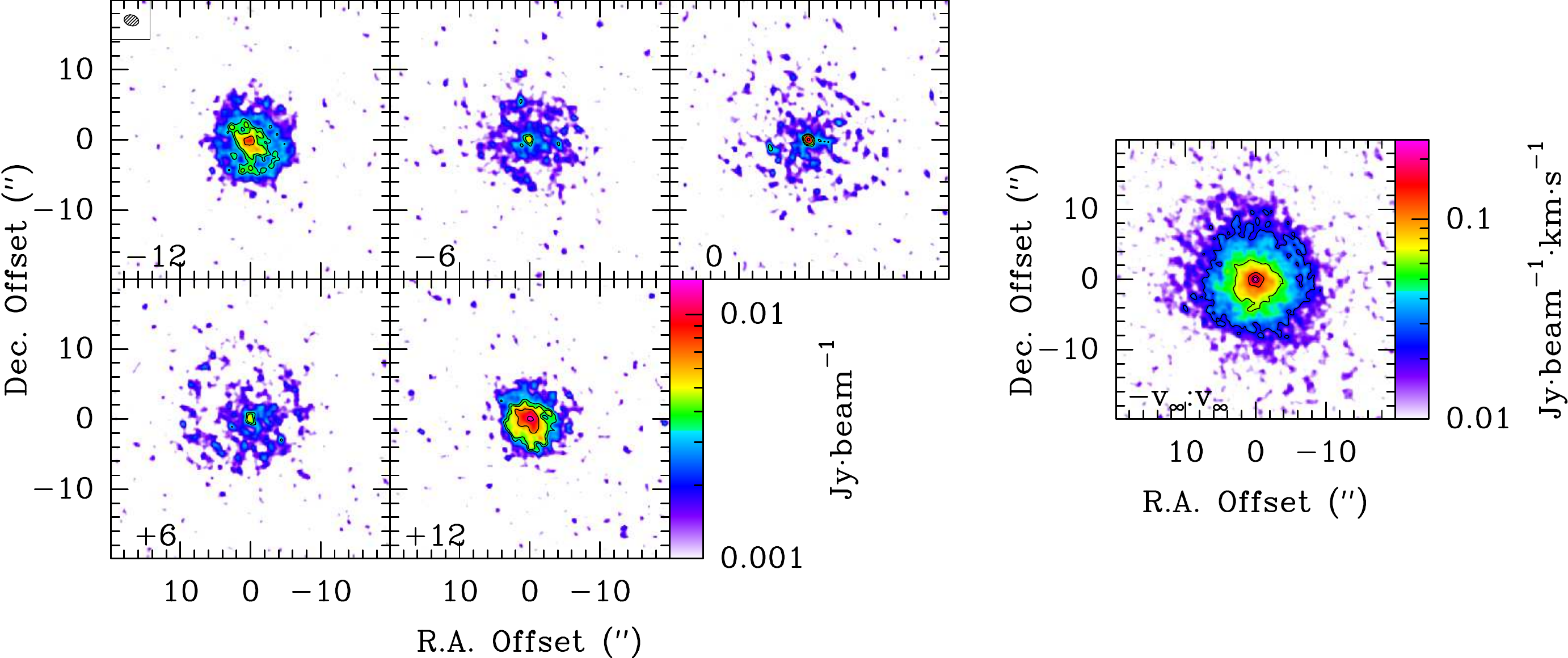}
\caption{As in Figure\,\ref{fig:29sio_2-1_maps}, but for $^{30}$SiS $J$=5--4.}
\label{fig:30sis_5-4_maps}
\end{figure*}

\begin{figure*}[h]
\centering
\includegraphics[scale=0.50]{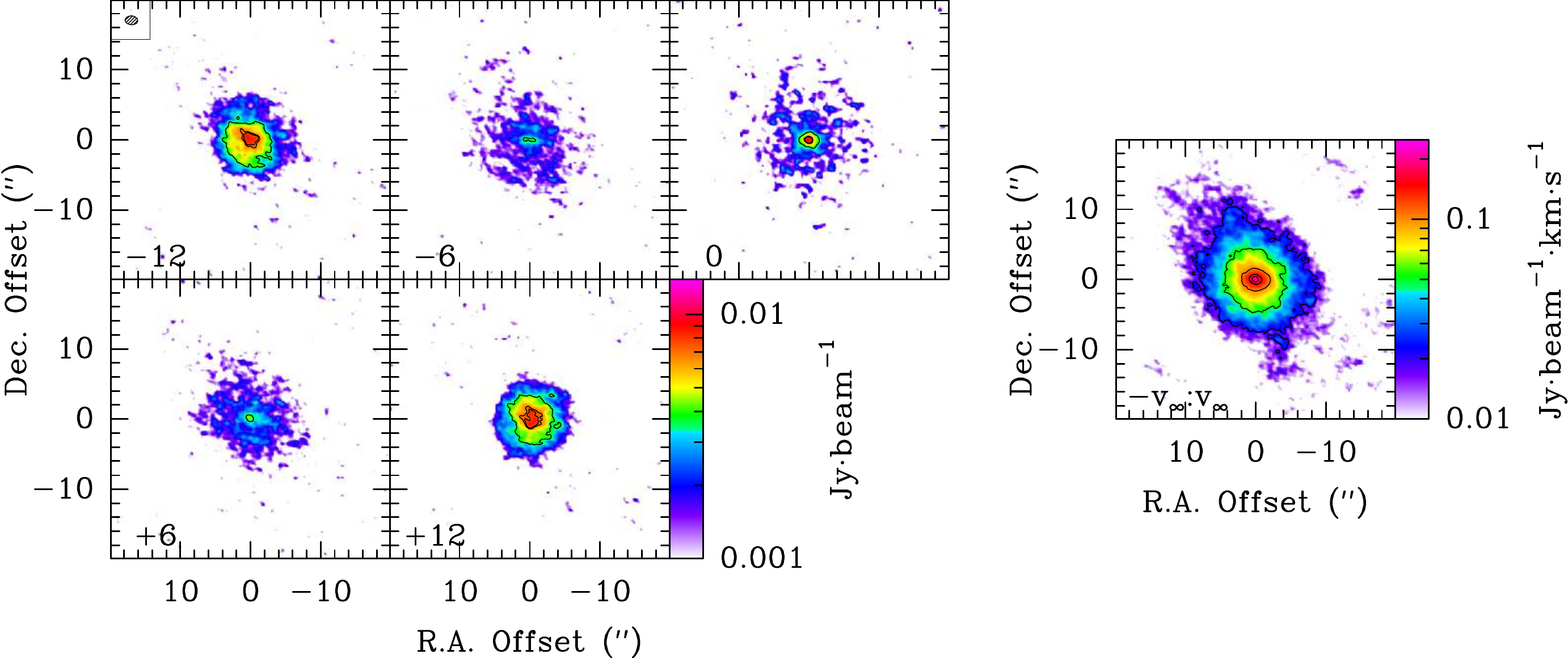}
\caption{As in Figure\,\ref{fig:29sio_2-1_maps}, but for $^{30}$SiS $J$=6--5.}
\label{fig:30sis_6-5_maps}
\end{figure*}

\begin{figure*}[h]
\centering
\includegraphics[scale=0.50]{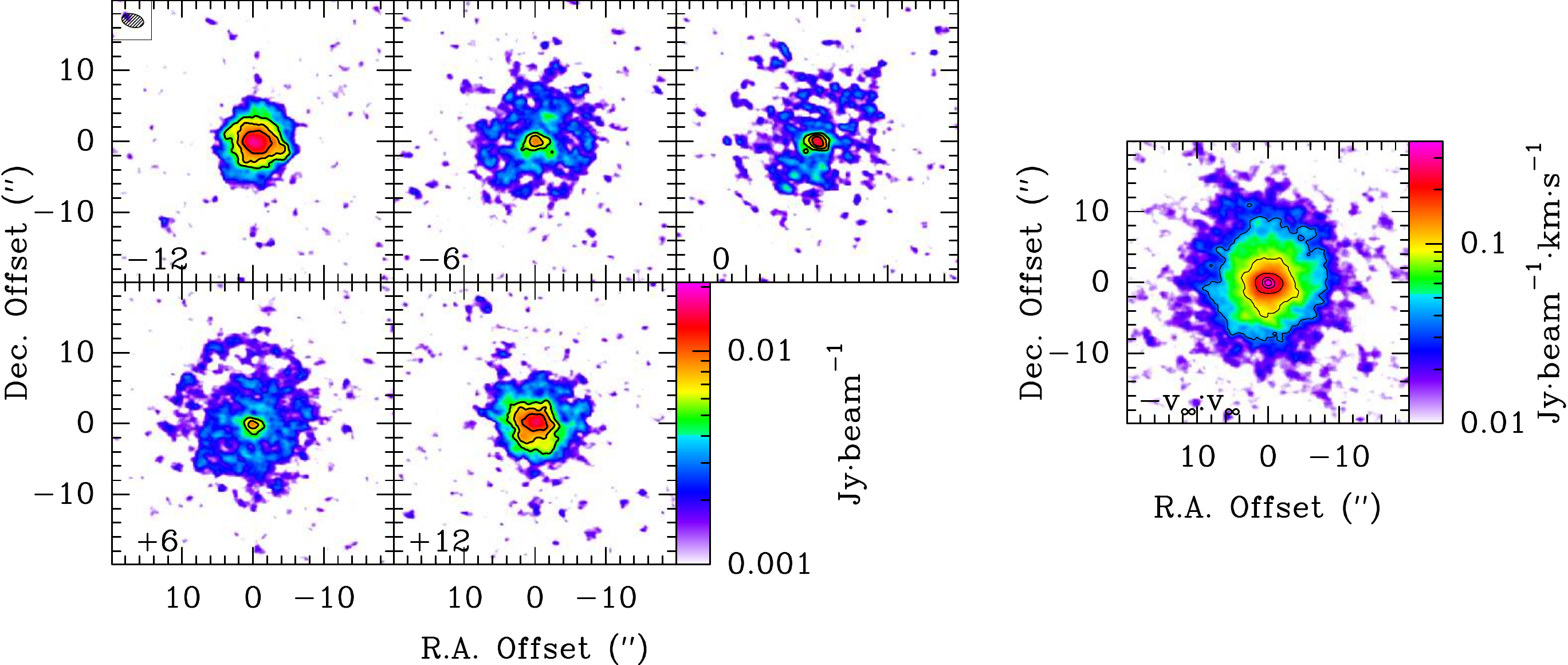}
\caption{As in Figure\,\ref{fig:29sio_2-1_maps}, but for Si$^{34}$S $J$=5--4.}
\label{fig:si34s_5-4_maps}
\end{figure*}

\begin{figure*}[h]
\centering
\includegraphics[scale=0.50]{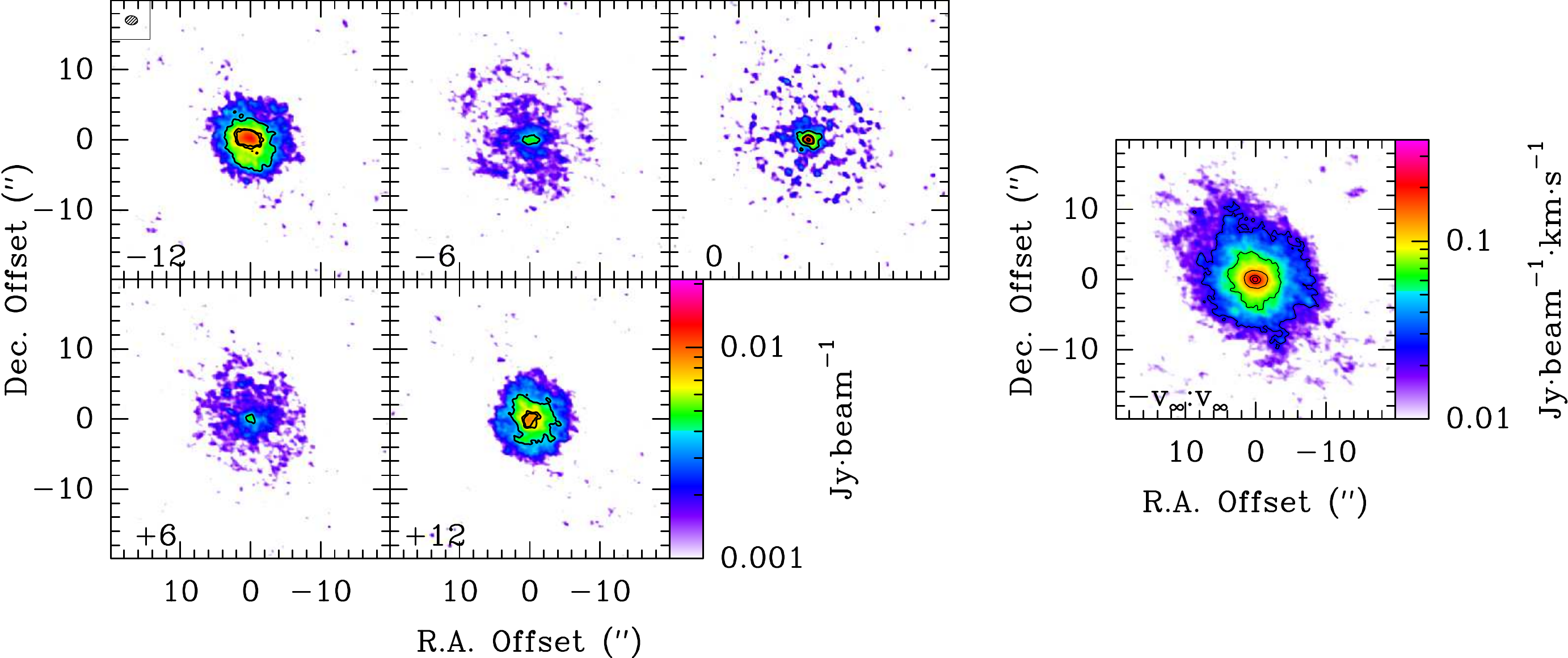}
\caption{As in Figure\,\ref{fig:29sio_2-1_maps}, but for Si$^{34}$S $J$=6--5.}
\label{fig:si34s_6-5_maps}
\end{figure*}

\begin{figure*}[h]
\centering
\includegraphics[scale=0.50]{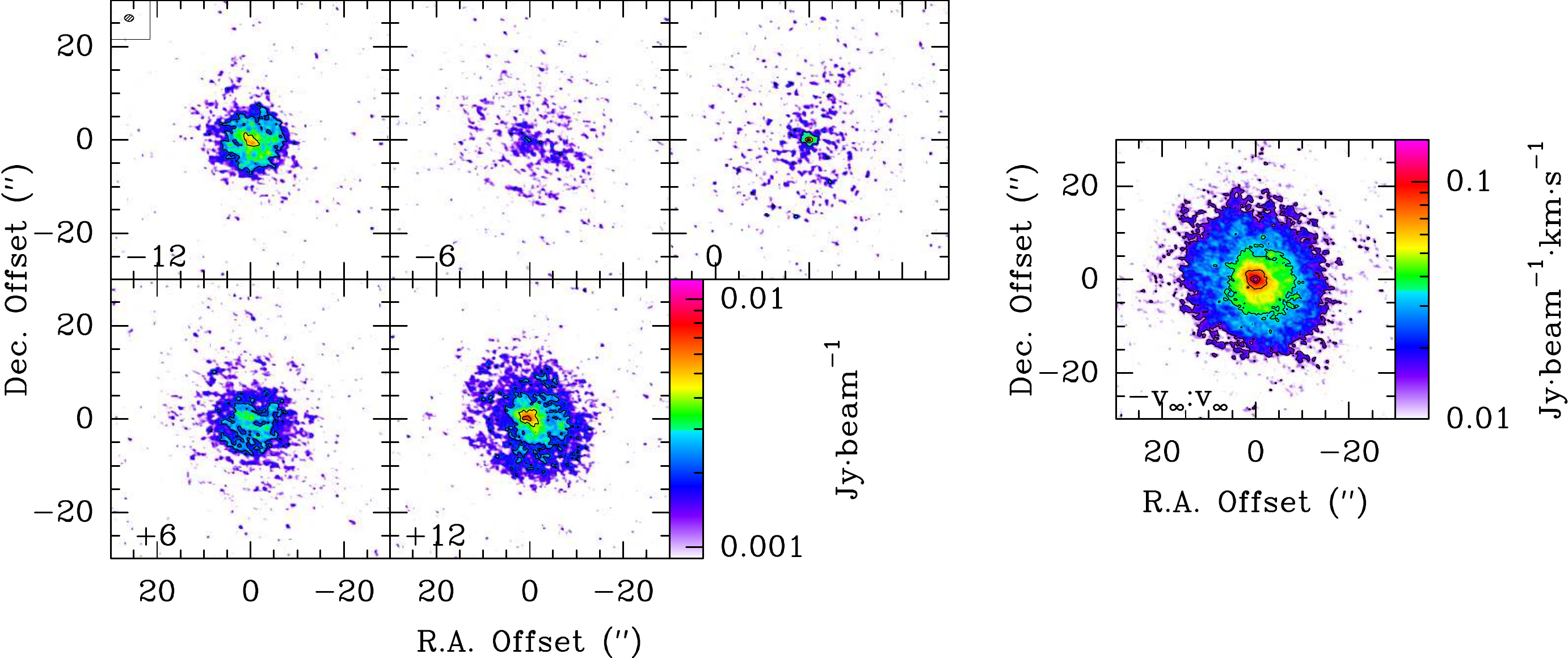}
\caption{As in Figure\,\ref{fig:29sio_2-1_maps}, but for $^{13}$CS $J$=2--1.}
\label{fig:13cs_2-1_maps}
\end{figure*}

\begin{figure*}[h]
\centering
\includegraphics[scale=0.50]{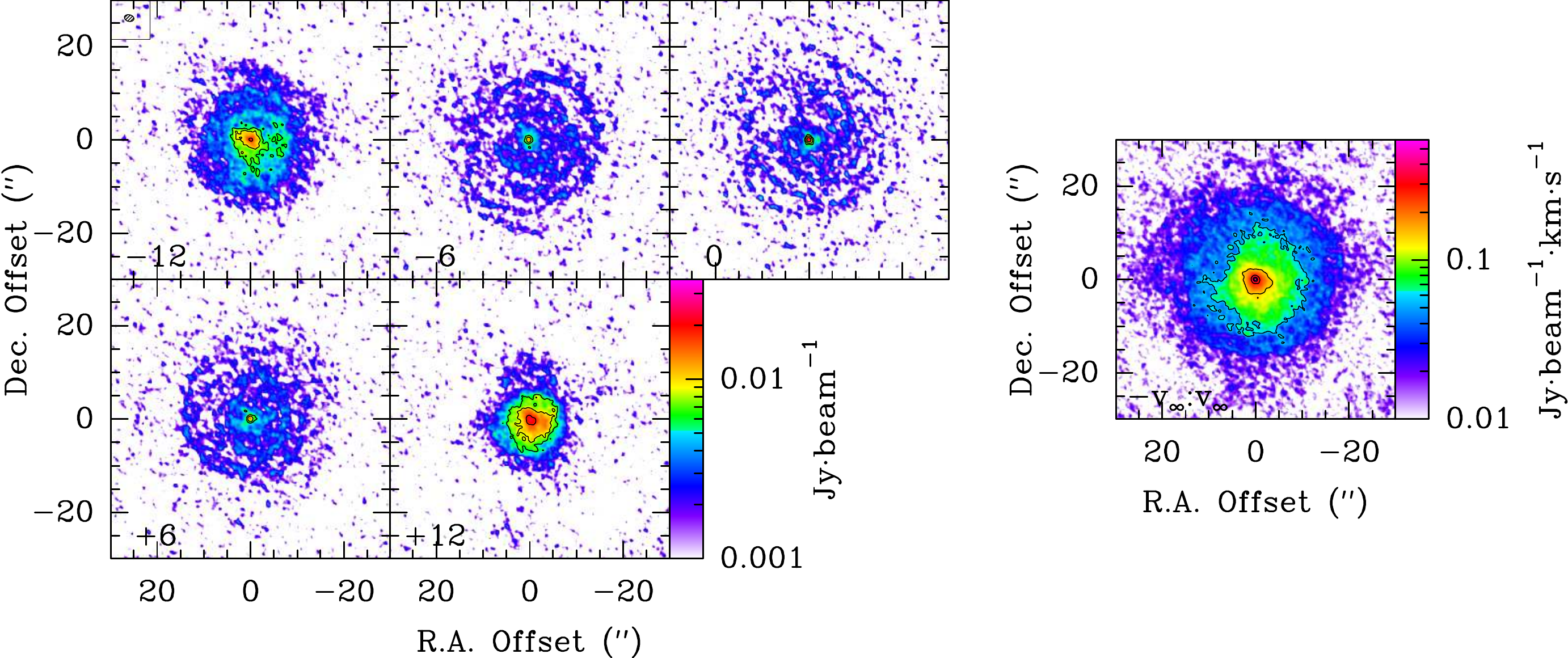}
\caption{As in Figure\,\ref{fig:29sio_2-1_maps}, but for C$^{34}$S $J$=2--1.}
\label{fig:c34s_2-1_maps}
\end{figure*}

\begin{figure*}[h]
\centering
\includegraphics[scale=0.50]{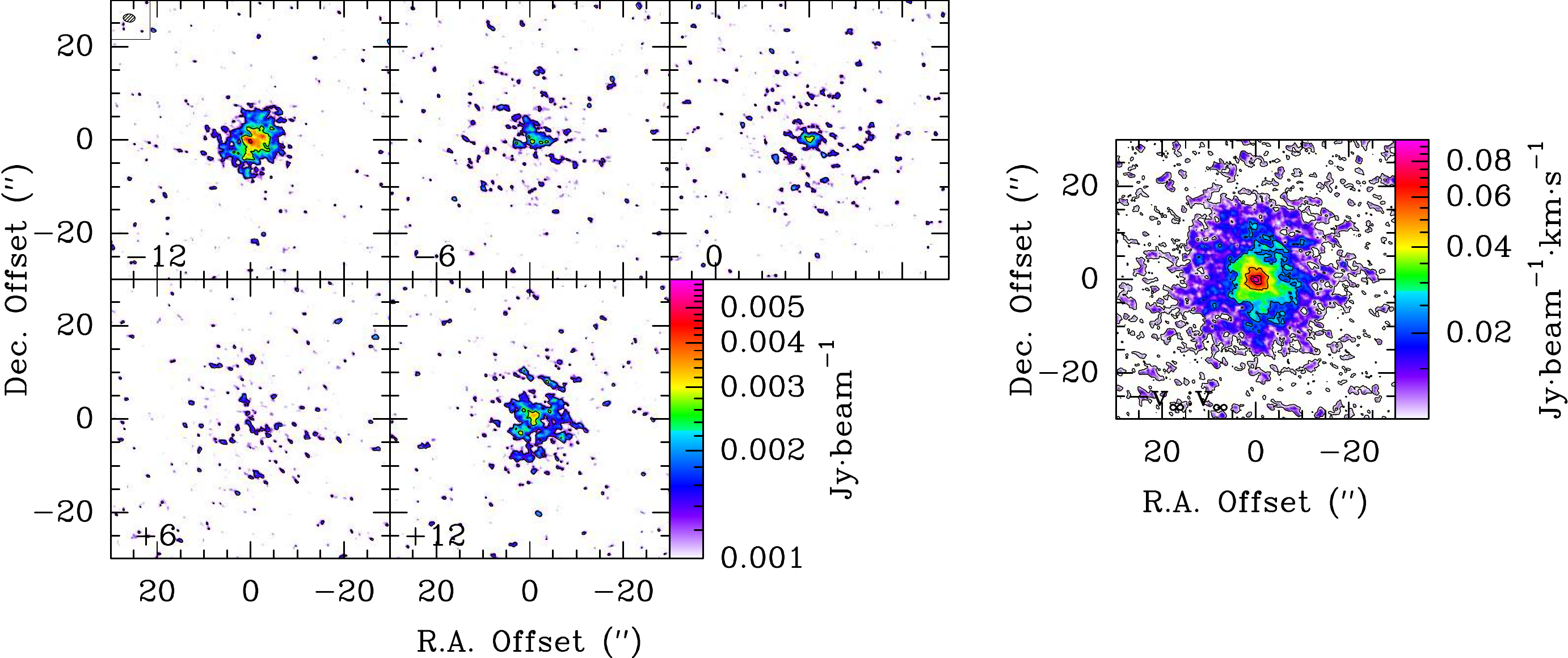}
\caption{As in Figure\,\ref{fig:29sio_2-1_maps}, but for C$^{33}$S $J$=2--1.}
\label{fig:c33s_2-1_maps}
\end{figure*}

\clearpage
\newpage

\section{PV diagrams for the rest of the isotopologues}\label{sec:app_iso_pvs}
In this section we present the PV diagrams for the rest of the detected lines of isotopologues:
the $J$=2--1 line of $^{29}$SiO and $^{30}$SiO; 
the $J$=5--4 and 6--5 lines of $^{29}$SiS, $^{30}$SiS, and Si$^{34}$S;
and the $J$=2--1 line of $^{13}$CS, C$^{34}$S, and C$^{33}$S. 

\begin{figure*}[h]
\centering
\includegraphics[scale=0.50]{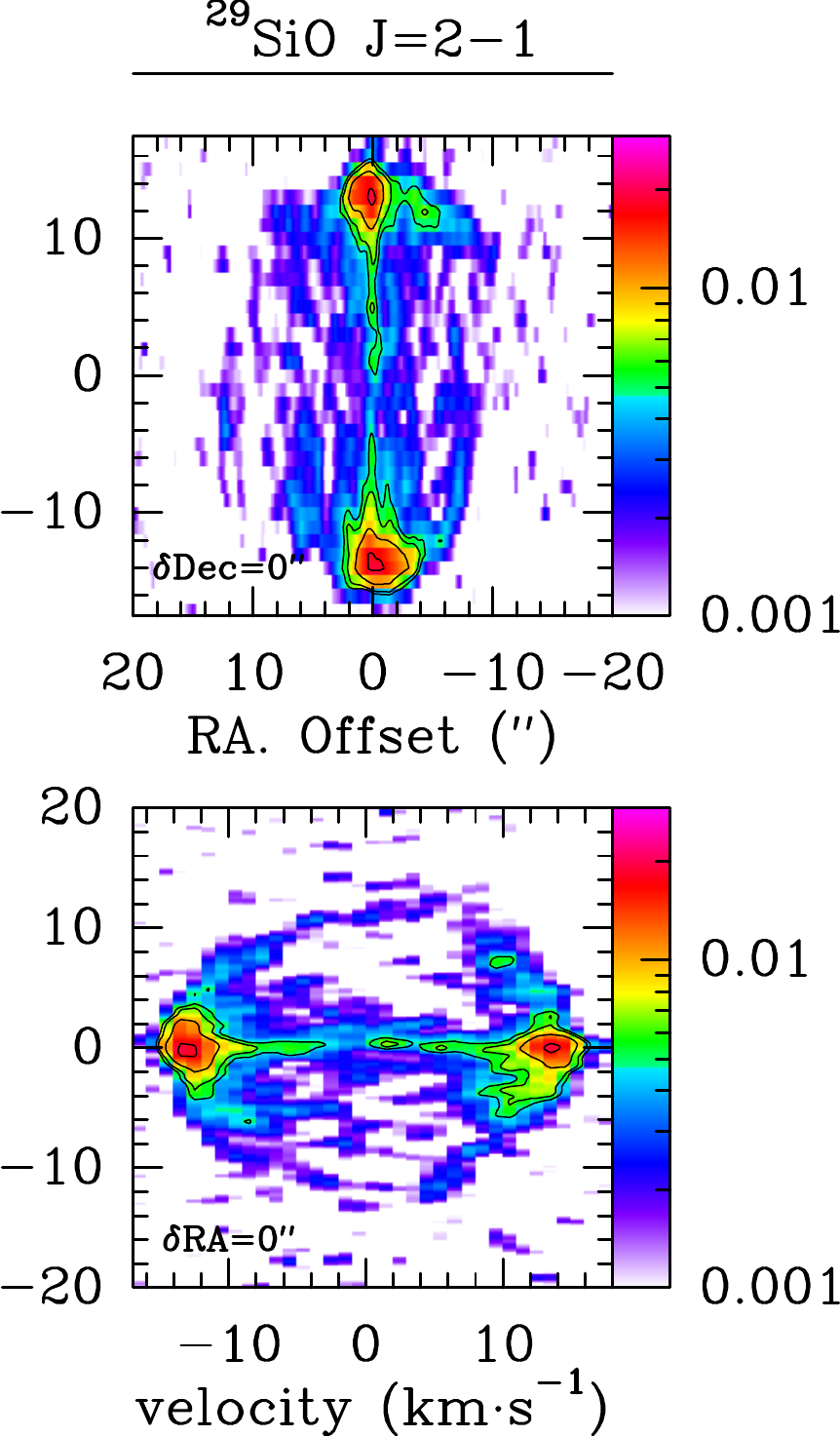}
\includegraphics[scale=0.50]{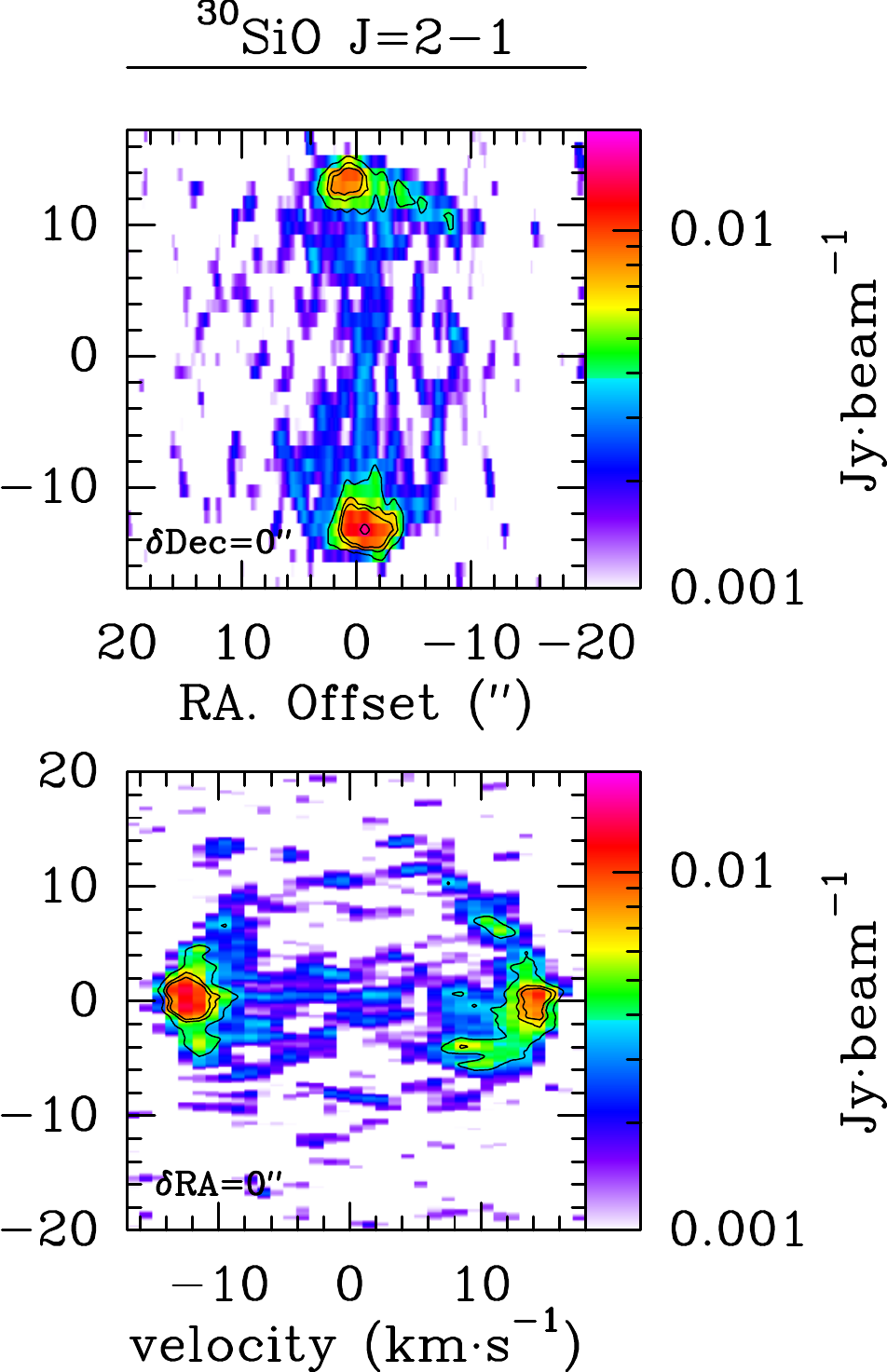}
\caption{PV diagrams for the detected lines of the rest of SiO isotopologues. 
The contours shown in black correspond to 5$\sigma$, 25\%, 50\%, and 90\%\ of the peak emission (see Table\,\ref{tab:summary}).
\textit{Top:} 
PV diagram of the flux density corresponding to a plane with a declination offset (see Section\,\ref{sec:obs}) equal to zero.
\textit{Bottom:} 
PV diagram of the flux density corresponding to a plane with a right ascension offset (see Section\,\ref{sec:obs}) equal to zero.
}
\label{fig:sio_iso_pvs}
\end{figure*}

\begin{figure*}[h]
\centering
\includegraphics[scale=0.50]{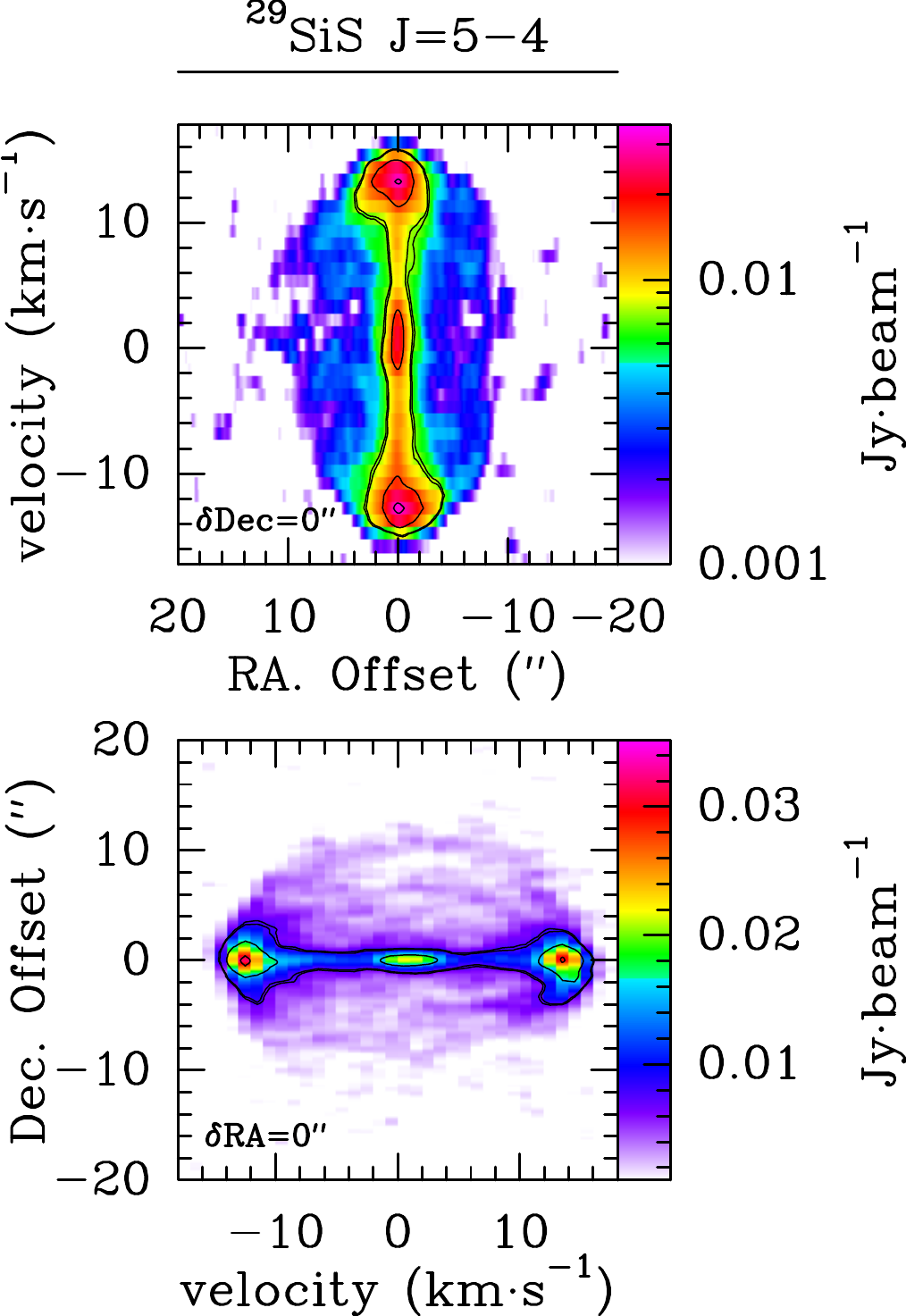}
\includegraphics[scale=0.50]{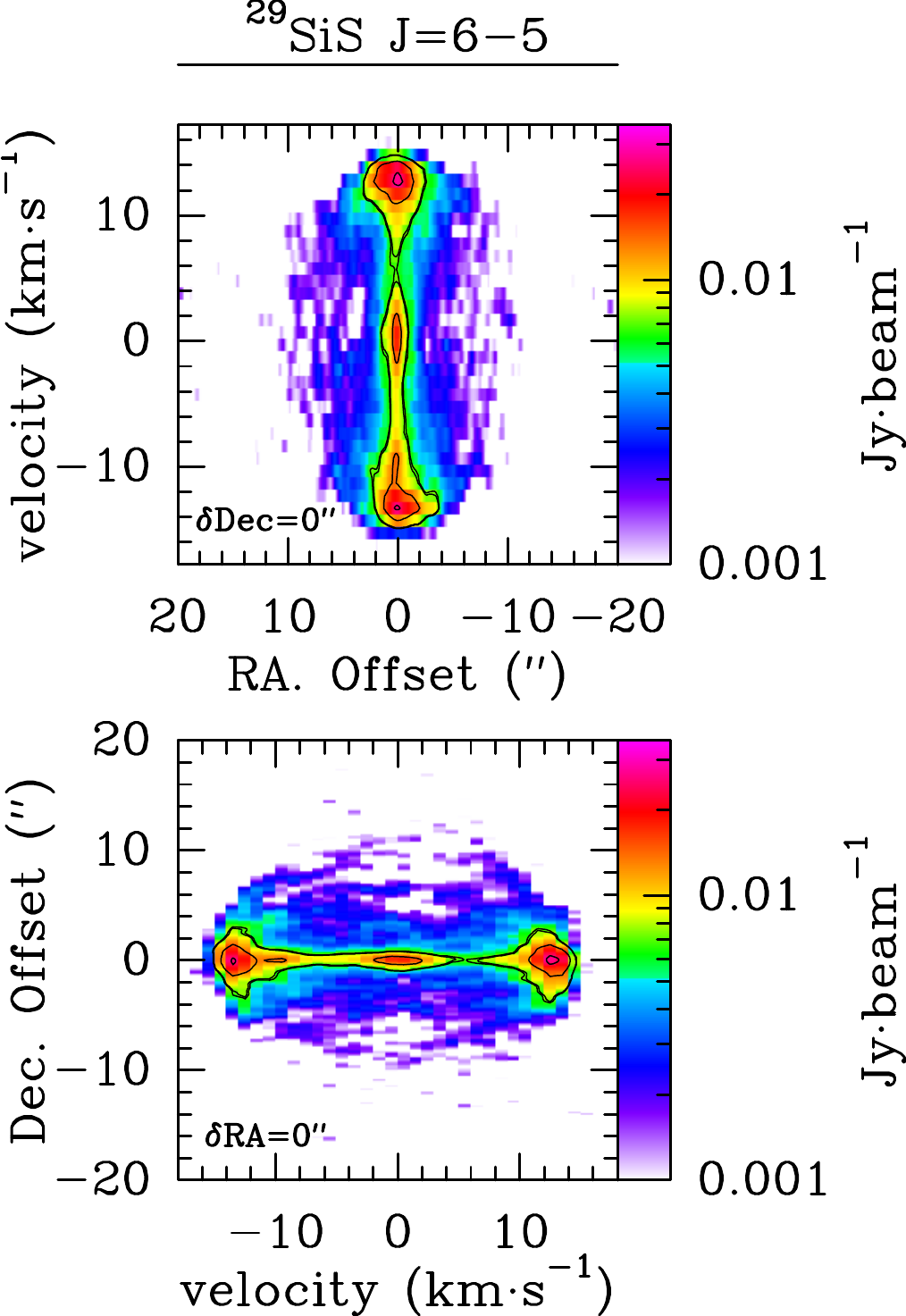}
\includegraphics[scale=0.50]{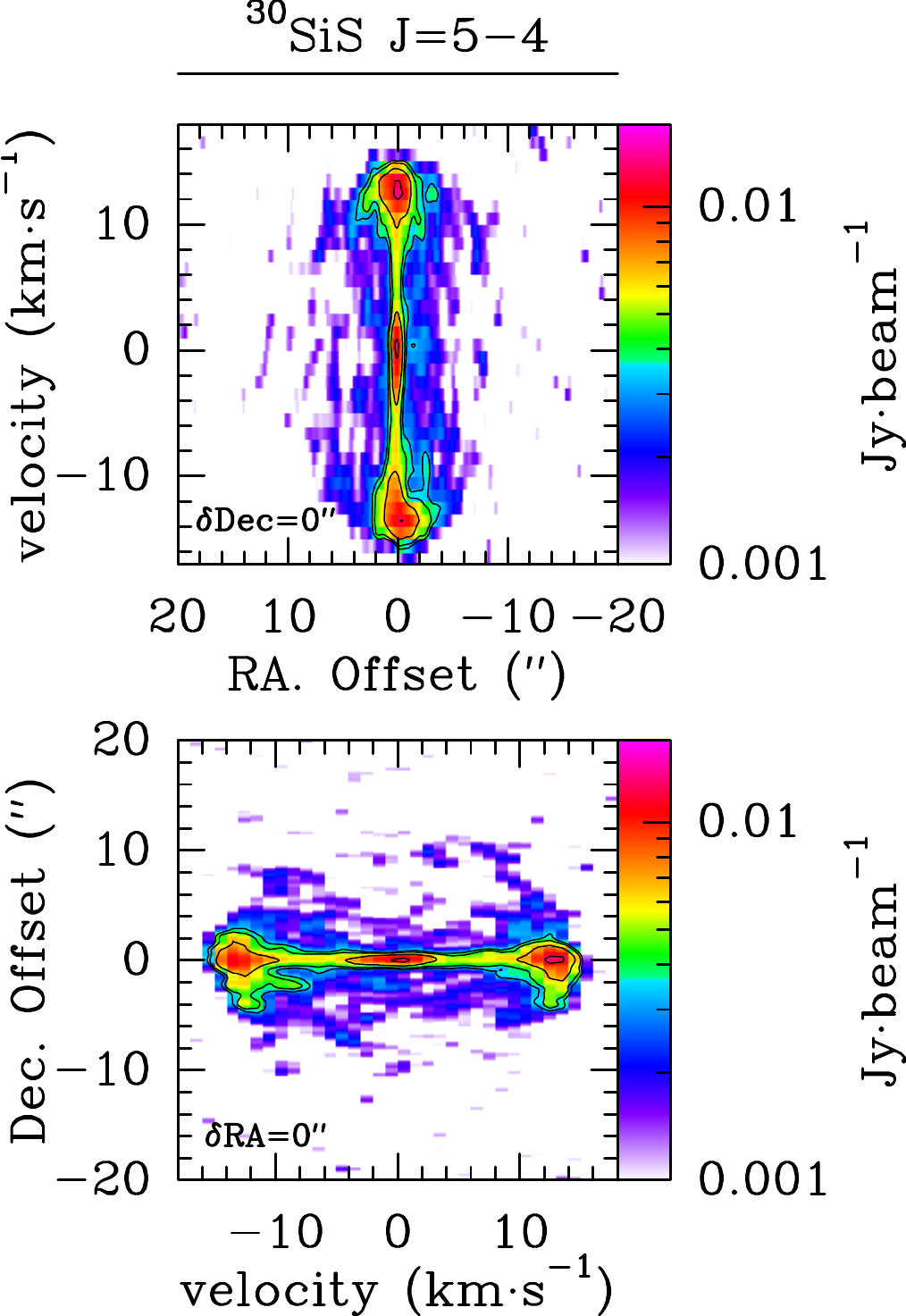}
\includegraphics[scale=0.50]{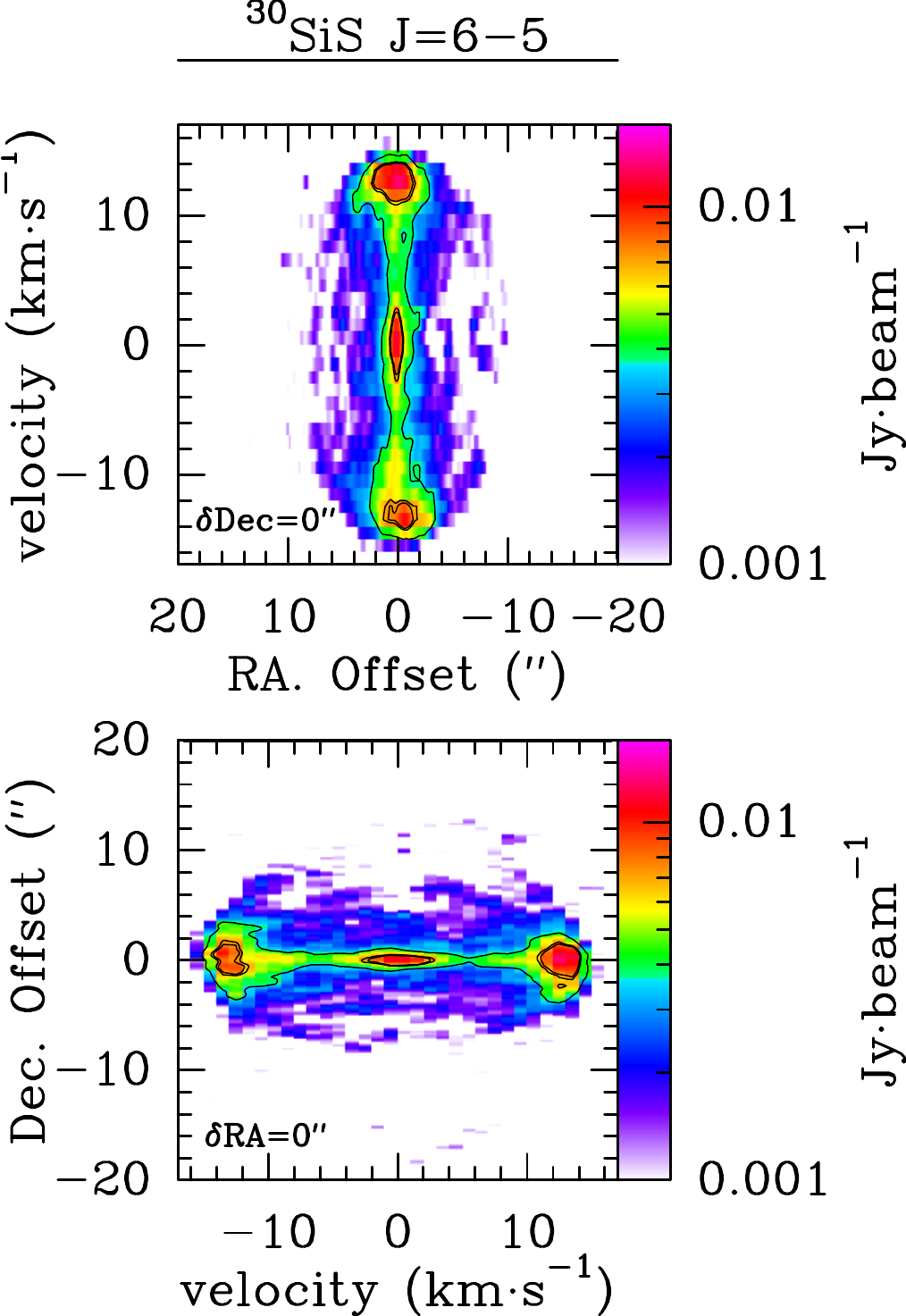}
\includegraphics[scale=0.50]{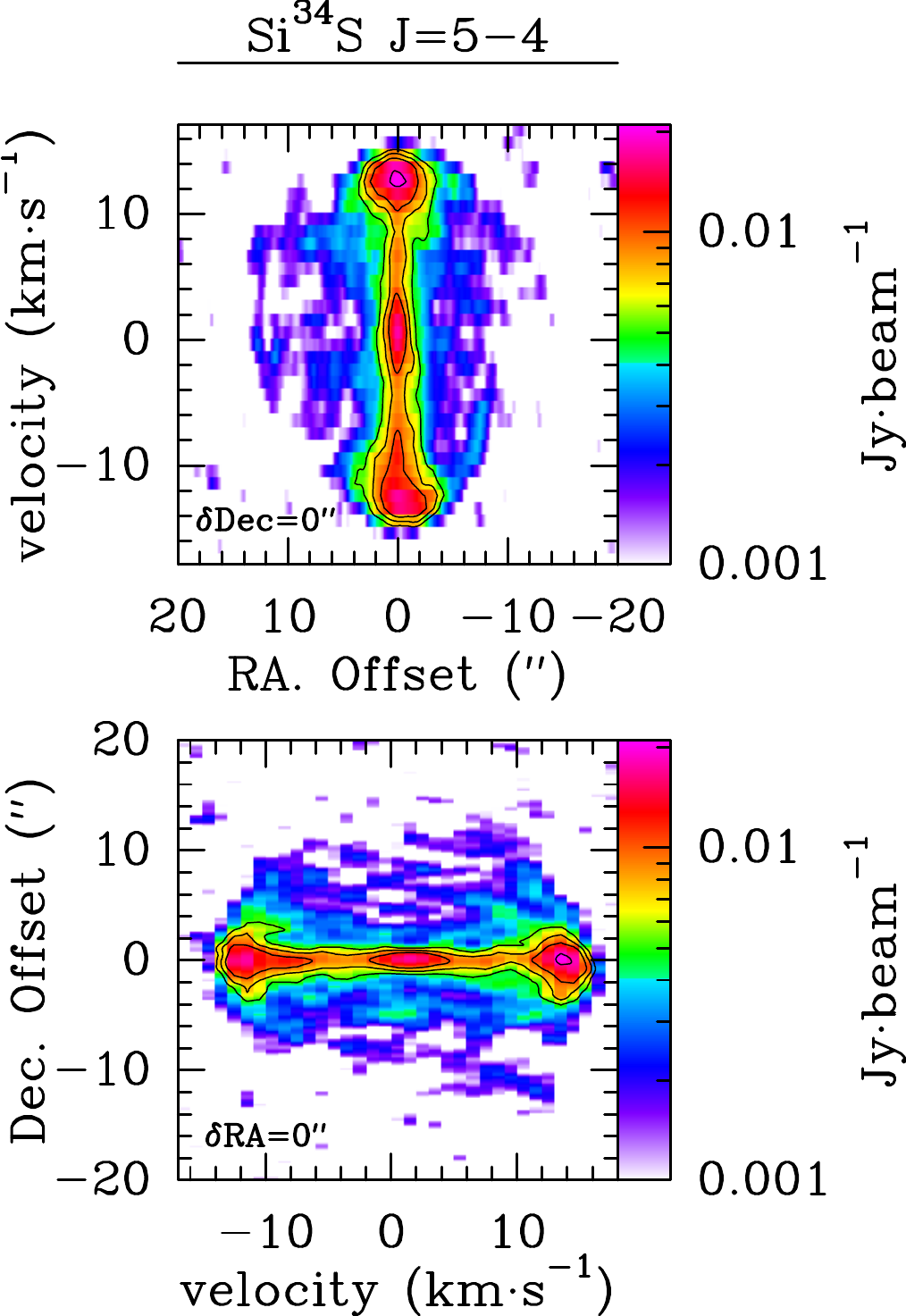}
\includegraphics[scale=0.50]{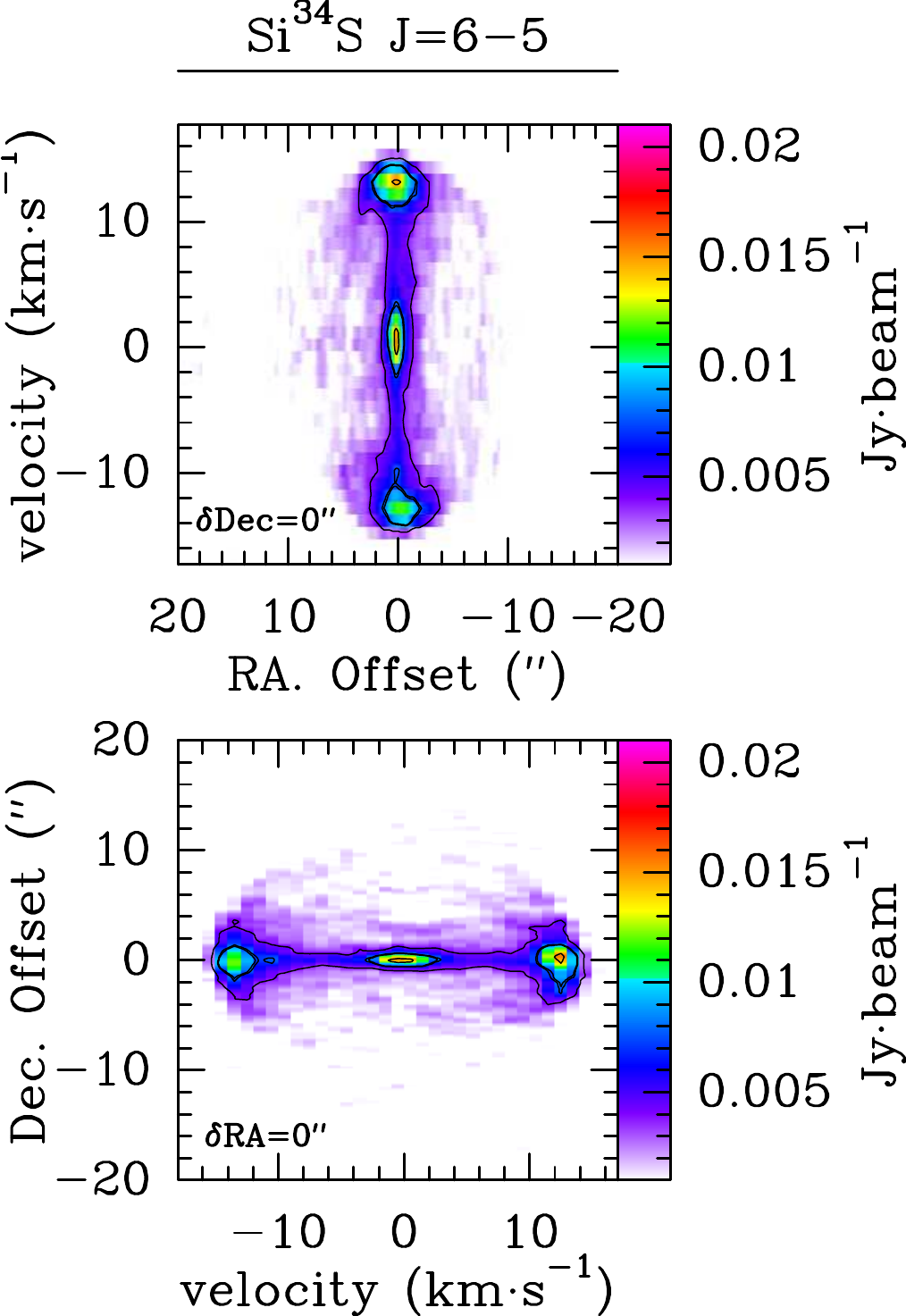}
\caption{As in Figure\,\ref{fig:sio_iso_pvs}, but for the SiS isotopologues}
\label{fig:sis_iso_pvs}
\end{figure*}

\begin{figure*}[h]
\centering
\includegraphics[scale=0.50]{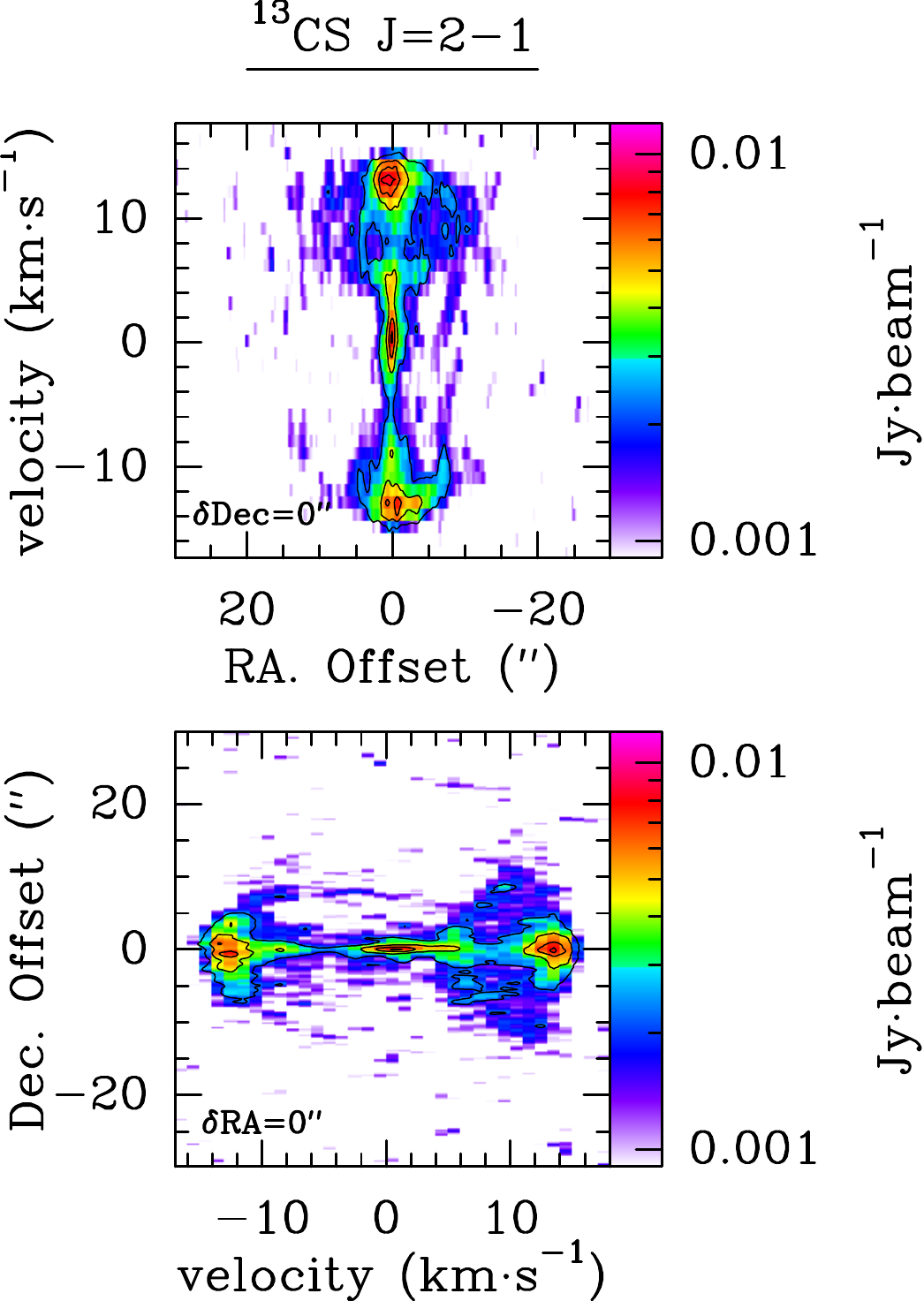}
\includegraphics[scale=0.50]{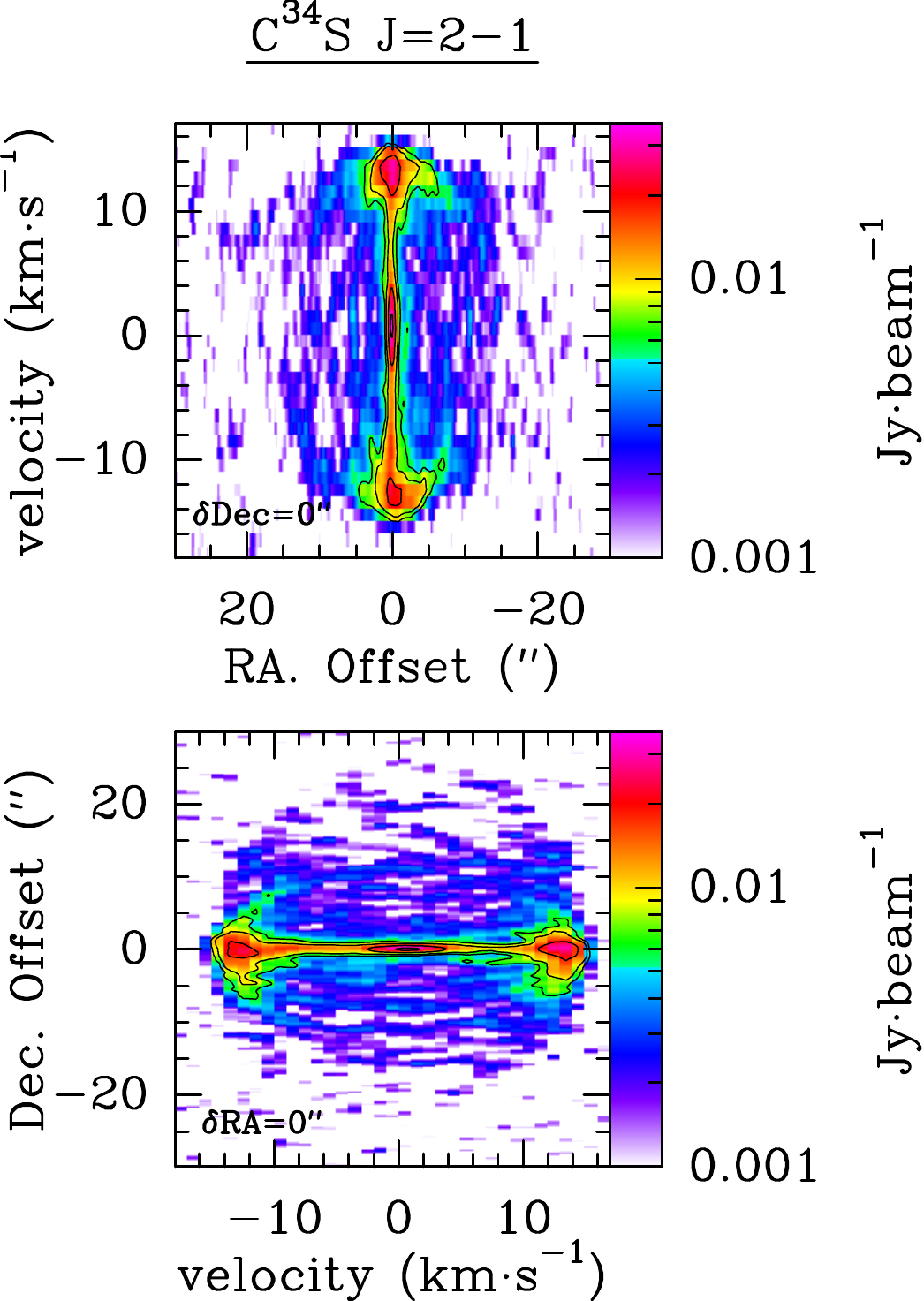}
\includegraphics[scale=0.50]{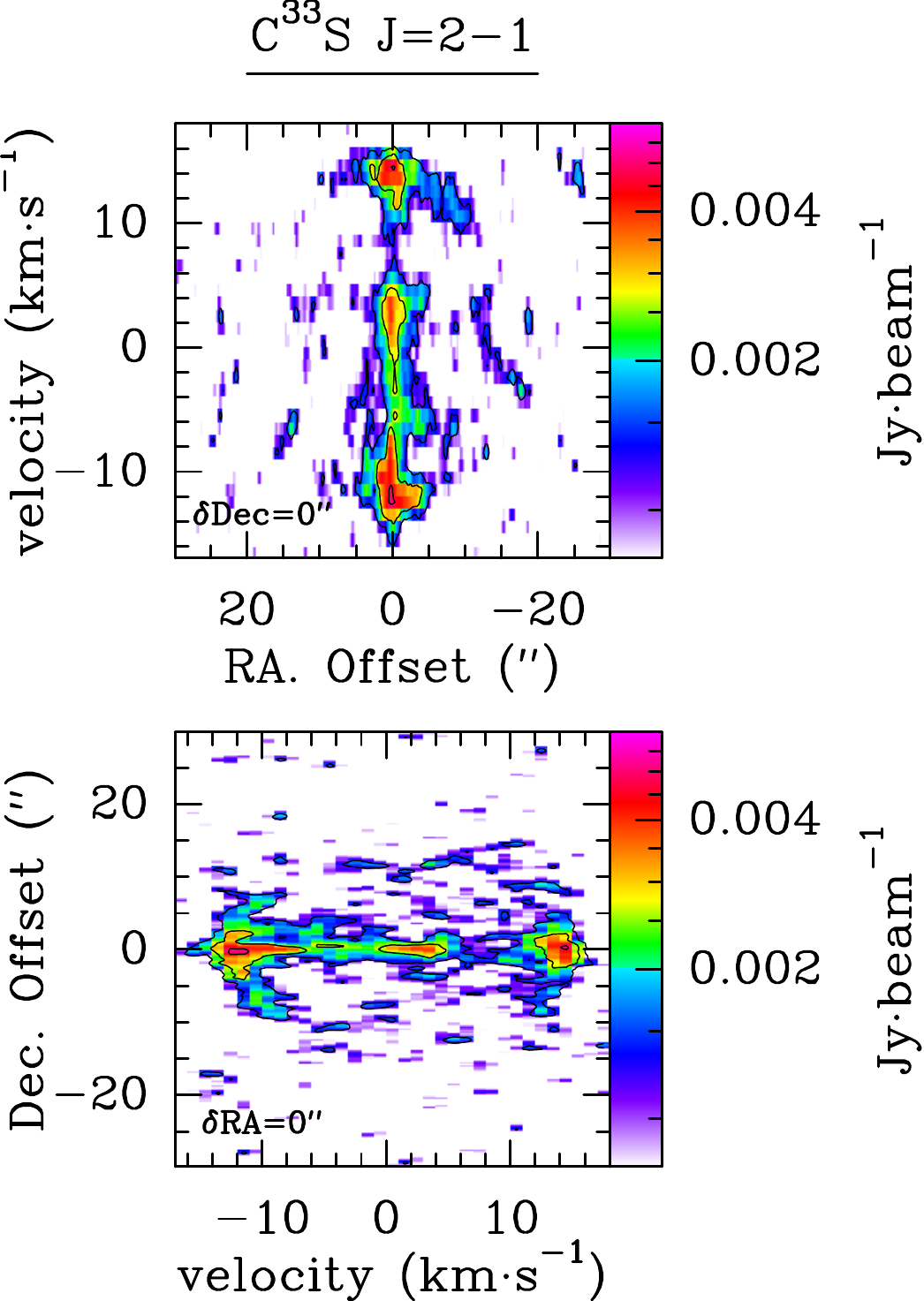}
\caption{As in Figure\,\ref{fig:sio_iso_pvs} but for the CS isotopologues}
\label{fig:cs_iso_pvs}
\end{figure*}

\end{document}